\documentclass[twocolumn,letterpaper,amsmath,amssymb]{revtex4}
\usepackage{graphicx}
\usepackage{dcolumn}
\usepackage{bm}
\begin{document}


%
%

\newcommand{\rf}[1]{(\ref{#1})}
\newcommand{\bfomega}{ \mbox{\boldmath{$\omega$}}}

\title{Granular Packings: 
Nonlinear elasticity, sound propagation
and collective relaxation dynamics}

\author{Hern\'an A. Makse$^1$, Nicolas Gland$^{2,1}$, David L.
Johnson$^3$, and Lawrence Schwartz$^3$}

\affiliation {
$^1$ Levich Institute and Physics Department, City College of
New York,
New York, NY 10031, US\\
$^2$ Ecole Normale Sup\'erieure, Departement T.A.O., 24 Rue
Lhomond, 75005,
Paris, France\\
$^3$ Schlumberger-Doll Research, Old Quarry Road, Ridgefield, CT
06877, US}

\date{Phys. Rev. E {\bf 70}, 061302 (2004)}

\begin{abstract}

\noindent

Experiments on isotropic compression of a granular assembly of spheres
show that the shear and bulk moduli vary with the confining pressure
faster than the $1/3$ power law predicted by Hertz-Mindlin effective
medium theories (EMT) of contact elasticity. Moreover, the ratio
between the moduli is found to be larger than the prediction of the
elastic theory by a constant value. The understanding of these
discrepancies has been a longstanding question in the field of
granular matter.  Here we perform a test of the applicability of
elasticity theory to granular materials. We perform sound propagation
experiments, numerical simulations and theoretical studies to
understand the elastic response of a deforming granular assembly of
soft spheres under isotropic loading.  Our results for the behavior of
the elastic moduli of the system agree very well with experiments.  We
show that the elasticity partially describes the experimental and
numerical results for a system under compressional loads. However, it
drastically fails for systems under shear perturbations, particularly
for packings without tangential forces and friction.  Our work
indicates that a correct treatment should include not only the purely
elastic response but also collective relaxation mechanisms related to
structural disorder and nonaffine motion of grains.

\end{abstract}
\maketitle


\section{Introduction and Objectives}
\label{intro}

The acoustic properties of granular porous materials confined by
an external stress can be extremely nonlinear as compared to
continuum elastic solids \cite{powders,goddard,guyer,hans0}.
Many industrial applications, such as the optimization of well
location in an oil reservoir, depend crucially on the correct
interpretation of nonlinear acoustic effects in granular
materials, as exemplified by the large variation of the sound
speeds or the elastic constants of the granular formation as a
function of the external stress \cite{sinha,footnote}.

Important insight into this problem comes first from the
Hertz-Mindlin contact theory to model the intergrain forces
\cite{johnson,landau}. In this case, nonlinearity arises from the
increase  with the external stress of the contact area between two
spherical grains. Conventional theories describing this problem in
the framework of elasticity of continuum media \cite{landau}
consider a uniform strain at all scales, and the displacement
field of the grains is affine with the macroscopic deformation
(the affine approximation).
Here, one computes the stresses in terms of the strains by
considering the disordered medium as an effective medium that
exerts a mean-field force (as given by contact Hertzian theory)
 on a single representative grain. This approximation is usually
referred to as the Effective Medium Theory (EMT)
\cite{mindlin2,digby,walton,johnson1}.

As shown in a short letter \cite{mgjs} and the studies of other
groups \cite{goddard}, the EMT does not successfully explain the
mechanical properties of cohesionless granular assemblies. The
main prediction of the theory is the scaling of the bulk modulus,
$K$, and shear modulus, $\mu$, with the pressure, $p$, as $K \sim
\mu \sim p^{1/3}$. However there is a large volume of experiments
for irregular sand grains as well as spherical glass beads which
show anomalous scaling characterized by exponents varying between
$1/3$ and $1/2$ (for a comprehensive review see Goddard
\cite{goddard} and for a review in the geotechnical literature see
\cite{woods}). Some studies have suggested that a $\sim p^{1/2}$
scaling is more appropriate for describing the nonlinear variation
of the moduli \cite{goddard}.

Here we extend the results of \cite{mgjs} and investigate the
applicability of elasticity theory  to granular matter
by means of experiments, computer simulations,
and analytical calculations. We first develop a series of acoustic
experiments to characterize the nonlinear elastic behavior of
non-cohesive dry granular materials under a wide range of external
pressures. From this experimental study we conclude that a
microscopic study is needed in order to elucidate the deficiencies
of existing granular theories. Then we perform a Molecular
Dynamics (MD)  simulation to give
microscopic insight into the relaxation mechanism of granular
materials. Finally, we offer a simple theory of relaxation going
beyond the effective medium approximation of elasticity.

We calculate the elastic moduli, $K$ and $\mu$,  of a disordered array of
elasto-frictional Hertz-Mindlin spherical grains.
Numerical simulations resolve the question as to whether the
problem lies in the treatment of intergrain contact or with the
EMT. We find agreement between our simulations and the
experiments, thus confirming the validity of the Hertz-Mindlin
contact theory to glass bead aggregates  composed of frictional
particles.

Regarding the anomalous  pressure dependence of the moduli, we
find that there are several nonlinearities which preclude the
proper definition of a scaling behavior as a function of pressure.
We find a regime at low pressure where the coordination number and
the volume fraction do not change much from their minimal values.
In this regime the $p^{1/3}$ scaling is approximately valid.
However, for pressures larger than 10 MPa the increase of the
coordination number and volume fraction induces other
nonlinearities and therefore no simple scaling behavior can be
defined.

In reality around 10 MPa there is crossover from a $p^{1/3}$ to a
$p^{5/9}$ scaling at larger pressures.  Thus, we conclude that the
scattering in the experimental values of the pressure exponent
might be explained by the fact that the exponent is actually
changing continuously from 1/3 to 5/9. This is especially true in
the regime where the experiments are usually done, near the
crossover region at 10 MPa.
Similar conclusions have been reached by Luding \cite{luding}
who also found that EMT needs to take into account 
the coordination number and its dependence on pressure.

Our most important results relate to the effect of friction and
stress relaxation on the behavior of the elastic constants.
Firstly, we find that the elastic formulation gives a reasonable
description for the response of the system to compressional loads,
i.e. the bulk modulus is reasonably well defined with the simple
EMT. However, our simulations establish that the EMT is inadequate
in describing the response of the material under shear
perturbations.

The numerical  simulations indicate that EMT fails because it does
not properly allow the grains to relax from the affine deformation
imposed by the external boundary. The affine
assumption is that, under an infinitesimal symmetric macroscopic
deformation, each grain translates according to the direction of
the macroscopic strain, and it does not rotate. Moreover, no
further relaxation mechanism is allowed. Such an homogeneous
strain field is consistent with the local force balance of grains
only in an ordered system.  For disordered packings an
inhomogeneous strain develops at the local level. After the
application of an affine strain the particles experience an
unbalanced force since they are not, in general, in a symmetric
environment. Consequently, the particles will move to a position
different to that predicted by the affine approximation, so that
the net force on each particle is zero. Similarly for torques and
rotations.

Here we show that the assumption of affinity is approximately
valid for the bulk modulus and seriously flawed for the shear
modulus. For this reason, the EMT prediction differs significantly
from the experimental value. Thus, the principal source of
deviation from EMT is the breakdown of the uniform strain
assumption.

To quantify the breakdown of EMT for the shear modulus we focus
our studies to two cases: frictionless grains interacting via only
normal forces (this system is said to be path-independent and it
is thought to describe compressible foams and emulsions) and
systems with elastic tangential forces and Coulomb frictional
forces (these systems are path-dependent and describe dry granular
materials).

The largest disagreement between theory and simulations is found
for frictionless systems; the difference is more pronounced at low
confining stress where we show that the system is in a state of
marginal rigidity at a minimal mean coordination number equal to 6
in 3D (or 4 in 2D). We find that after the application of an
external shear strain there is a nearly complete relaxation of the
system to the applied shear; a result that cannot be captured
under the framework of elasticity.  We show that a new scaling
behavior with pressure might describe the data for frictionless
particles better: $\mu(p) \sim p^{2/3}$ as $p\to 0$.

We interpret this result in the framework of critical phenomena:
as $p \to 0$ the system approaches a critical point at a mean 
coordination number $Z_c=2D$ in $D$ dimensions, and a volume
fraction of random close packing $\phi_c \approx 0.64$. This point
describes a rigidity threshold state or a critical state of the
packing as defined by Alexander \cite{alexander} and it is where
the system  becomes ``isostatic''
\cite{isostatic,edwards-grinev,ball}. The elastic moduli vanish
as a power-law of the pressure or volume fraction.  For any finite
pressure rigidity is achieved, since $Z>Z_c$. Near the rigidity
threshold the reference packing structure has a power law
dependence on the pressure, modifying the scaling of $\mu(p)$
predicted by the Hertz theory.

When  friction and tangential elasticity is restored at the
intergrain contacts, the agreement between theory and simulations
(and in this case experiments) improves with respect to the
frictionless case.  This is because the existence of tangential
restoring forces reduces the extent to which the grains relax from
the assumed affine configuration.  Thus, the EMT provides a better
agreement with simulations and with experiments for frictional
grains than for frictionless grains, but serious disagreements
still persist as we shall demonstrate.

We conclude that in order to develop a better understanding of the
problem, one must abandon the purely elastic framework and
consider granular matter as a full viscoelastic body. Collective
relaxation effects can account for the discrepancy in the shear
modulus in comparison with the elastic prediction: the corrections
increase dramatically in the case of loose materials and for
frictionless packings. A theory of single-particle relaxation is
offered as a first step in this direction. We also discuss our
results in the framework of recent theories of marginal rigidity,
jamming, melting and fragile matter.

{\it Applications.--} Part of the motivation for this research
derives from the fact that acoustics and nonlinear elastic logging
methods are at the forefront of the evolving technology to help
plan and optimize well location in oil exploration
\cite{oilfield}. In order to position a well correctly, the
knowledge of the stress distribution around the borehole is
essential. Mechanical properties of the granular formation
obtained from sonic logging can help predict formation strength,
while stress magnitude derived from sonic measurements helps in
predicting sanding problems in unconsolidated formations. Acoustic
measurements in granular materials provide the natural way to
understand the distribution of stress around the borehole.


Quite apart from the relevance to borehole logging, within the
field of seismic tomography there is a growing interest in
developing techniques to generate images of dissipation, along
with the more traditional images of impedance contrast.  This
extended seismic tomography would have impact, not only on the
understanding of hydrocarbon reservoirs but also e.g. on
techniques to monitor the migration of ground water pollutants.
This work represents an attempt to understand some of the
mechanisms of attenuation as well as the stress dependence of
sound velocities in granular materials.

The outline of the paper is as follows. Section \ref{back} reviews
the background of the problem of Effective Medium theories and
numerical approaches: MD simulations and intergrain forces for
granular materials, compressed emulsions and foams. Section
\ref{experiments} describes the experiments on sound propagation.
Section \ref{numerics} describes the numerical results and Section
\ref{sectheory} the theoretical results. We conclude in Section
\ref{summary} with a final outlook.

\section{Background}
\label{back}

The problem of elastic properties of granular materials has been
treated by many researchers since the pioneering work of Mindlin
in the 50's \cite{goddard,mindlin2,domenico,yin,winkler,ishibashi,liu1,jussieu,pgg,jenkins,nesterenko,hascoet}.
However, a general solution to this problem is still lacking.

In a typical experiment, a set of cohesionless glass beads is
confined at a hydrostatic stress, $p$, and the compressional sound
speed, $v_p$, and the shear sound speed, $v_s$, are measured as
functions of stress (see for instance Domenico \cite{domenico},
Yin \cite{yin},
 and \cite{goddard,winkler,jussieu}).
The P-wave and S-wave speeds are related to the elastic constants
of the material in the long-wavelength limit:
\begin{equation}
v_p= 
\sqrt{\frac{K+4/3\mu}{\rho}}, \label{vp}
\end{equation}
\begin{equation}
v_s= \sqrt{\frac{\mu}{\rho}}, \label{vs}
\end{equation}
where $\rho$ is the mass density of the system.

\subsection{Contact mechanics.}
\label{background-forces}

In his seminal paper ``On the Contact of Elastic Solids'' H. Hertz
\cite{johnson,landau} used linear elasticity of continuum media to
calculate the normal force of two perfectly elastic spheres
pressed into contact considering no attraction or stickiness.
Hertz showed that two spherical grains in contact with radii $R_1$
and $R_2$ interact with a normal repulsive force

\begin{equation} F_n = \frac{2}{3}~ k_n R^{1/2}\xi ^{3/2}, \label{fn}
\end{equation}
where $R=2 R_1 R_2/(R_1+R_2)$, the normal overlap is $\xi=
(1/2)[(R_1+R_2) - |\vec{x}_1 - \vec{x}_2|]>0$, and $\vec{x}_1$,
$\vec{x}_2$ are the positions of the grain centers.  The normal
force acts only in compression, $F_n = 0$ when $\xi<0$.
The effective stiffness $k_n=4 \mu_g / (1-\nu_g)$ is defined in
terms of the shear modulus of the grains $\mu_g$ and the Poisson
ratio $\nu_g$ of the material from which the grains are made
(typically $\mu_g=29$ GPa and $\nu_g = 0.2$, for spherical glass
beads).

The situation in the  presence of a tangential force, $F_t$, is
more complicated.  In the case of spheres under oblique loading,
the tangential contact force was first calculated by Mindlin
\cite{mindlin}.
A general loading history can be described by the incremental
change in the tangential force $\Delta F_t$ and in the normal
force $\Delta F_n$. For the special case where the partial
increments do not involve microslip at the contact surface (i.e.,
$|\Delta F_t| < \mu_f \Delta F_n$, where $\mu_f$ is the kinematic
friction coefficient between the spheres, typically $\mu_f=0.3$)
Mindlin \cite{mindlin} showed that the tangential force is
\begin{equation}
 \Delta F_t= k_t (R \xi)^{1/2} \Delta s,
\label{ft} \end{equation} where $k_t = 8 \mu_g / (2-\nu_g)$, and
the variable $s$ is defined such that the relative shear
displacement between the two grain centers is $2s$.  This is
called the Mindlin ``no-slip'' solution. [See Appendix A for a
more general solution]. 

The incremental form Eq. (\ref{ft}) is
needed since the numerical value of the tangential force depends
upon the trajectory taken in the space ($\xi, s$), see
\cite{johnson1} for details. The tangential force is obtained by
integrating over the path taken by the spheres in contact subject
to the initial conditions: $F_n=0$, $F_t=0$ at $\xi=0, s=0$,
yielding:

\begin{equation}
F_t = \int_{path}  k_t (R \xi)^{1/2} ds. \label{Ft}
\end{equation}

Thus, a granular system with tangential elastic forces is said to
be path-dependent. By path dependency we mean that the work done
in deforming the system depends upon whether one first compresses
the system, then shears it, or first shears it then compresses.
The results depend upon the path taken and not just the
instantaneous final state. On the other hand, a system of spheres
interacting only via normal forces, Eq. (\ref{fn}), is said to be
path-independent, and the work does not depend on the way the
strain is applied.

As the shear displacement increases, the elastic tangential force
$F_t$ reaches its limiting value given by Amontons' law for no
adhesion, $F_t \leq \mu_f F_n$. Amontons' law (a special case of
Coulomb's law) adds a second source of path-dependency as well as
hysteresis to the problem.


\subsection{Effective medium theories (EMT) of granular elasticity}

The basic idea of elastic theories relevant to our study is that
the macroscopic work done in deforming the system is set equal to
the sum of the work done on each grain-grain contact and that the
latter is replaced by a suitable average
\cite{digby,walton,johnson1}. These theories are usually referred
to as the effective medium theory (EMT) and are based on
Hertz-Mindlin contact mechanics. In the case of an isotropic
deformable solid (for simplicity we describe the isotropic case),
the strain energy density per unit volume as a function of the
strain $\epsilon_{ij}$ is:
\begin{equation}
U(\epsilon_{ij}) =U_0 - p \epsilon_{ll} +  \mu_e (\epsilon_{ij} -
\frac{1}{3} \delta_{ij} \epsilon_{ll})^2 + \frac{1}{2}K_e
(\epsilon_{ll})^2 +O(\epsilon_{ij}^3)\:\:\:. \label{energy}
\end{equation}
This equation is equivalent to the expression
\begin{equation}
\sigma_{ij} = K_e \epsilon_{ll} \delta_{ij} + 2 \mu_e (\epsilon_{ij} -
\frac{1}{3} \delta_{ij} \epsilon_{ll}),
\label{stress-strain}
\end{equation} 
which determines the stress tensor $\sigma_{ij}$ in terms
of the strain tensor for an isotropic body \cite{landau}.
Here  
\begin{equation}
\epsilon_{ij} = 1/2 (\partial u_i/\partial x_j +
\partial u_j/\partial x_i),
\end{equation}
and the deviations  $u_i = x_i - R_i$
of the positions of the $N$ particles in the system, {\bf
\{x$_1,\ldots,$ x$_N$\} }, are measured from a suitable rigid
reference state
\begin{equation}
\mbox { \bf{\{R\} = \{R$_1$,\ldots, R$_N$\} } }, \label{R}
\end{equation}
around which one can expand consistently. This reference state is
straightforward for simple periodic systems. However, it is assumed
that this
expansion is also possible for amorphous solids with reference
states which are random, like in a packing of grains (see
Alexander \cite{alexander} for extensive discussions). The
subscript in $K_e$ and $\mu_e$ denotes that the values of the
moduli are calculated considering granular materials as purely
elastic solids. There are two assumptions inherent to the elastic
EMT:

\begin{itemize}
\item  All the spheres
are statistically the same, and it is assumed that there is an
isotropic distribution of contacts around a given sphere.

\item
 An
affine approximation is used, i.e., the spheres at position
$x_{j}$ are moved a distance $\delta u_i$ in a time interval
$\delta t$ according to the macroscopic strain rate
$\dot{\epsilon}_{ij}$ by
\begin{equation}
\delta u_i=\dot{\epsilon}_{ij} x_{j} \delta t. \label{affine}
\end{equation}
The grains are always at equilibrium due to the assumption of an
isotropic distribution of contacts, and further relaxation is not
required. This sort of mean field theory is analogous to a simple
average of non-linear spring constants.
\end{itemize}

It is important to notice that a definition of uniform strain
field is possible only under the  mean field approximation. This
assumption is also trivially correct for ordered packings.
However, for disordered systems, the affine approximation is
inconsistent with the local equilibrium of grains \cite{green3}.
We will come back to this crucial point later on.

With the above assumptions, the elastic energy Eq. (\ref{energy})
is set equal to a suitable average over the contacts, viz:

\begin{equation}
U = \frac{1}{V} \sum_{contact} \int \mbox{\bf F $\cdot$  $d$u}
\approx \frac{Z \phi}{V_0} \langle \int  \mbox{\bf F $\cdot$ $d$u}
\rangle,
\end{equation}
with $\mbox{\bf F} \cdot d\mbox{\bf u}= F_n d\xi + \mbox{\bf F}_t
\cdot d\mbox{\bf s}$, $Z$ is the average coordination number
defined as the average number of contacts per particle, $\phi$ is
the volume fraction of the sample, and $V_0$ is the volume of a
single grain. The EMT predictions for the bulk and shear modulus
for an isotropic system compressed at pressure $p$ are the
following:

We distinguish between two different models:

{ (1)\it  Path-independent models, $k_t=0$, frictionless grains}:
We consider that there is perfect slippage at the intergrain
contact. This corresponds to $F_t = 0$, only normal forces between
the particles. This case corresponds to path-independent forces,
and allows the use of an energy density function Eq.
(\ref{energy}) which depends only on the instantaneous position of
the particles. This case could be considered conservative since
the total work on a closed path is zero. A system of frictionless
spherical particles could be thought of as a model of compressed
emulsions and foams which are usually modeled as viscoelastic
spheres without tangential forces \cite{lacasse,durian,faraday}
(see Appendix C).

For the case of frictionless grains one finds:

\begin{equation}
K_e(p)= \frac{k_n}{12 \pi} ~ (\phi Z)^{2/3} ~\left(\frac{6 \pi
p}{k_n}\right)^{1/3}, \label{emt_k}
\end{equation}
\begin{equation}
\mu_e(p)= \frac{k_n}{20 \pi}~  (\phi Z)^{2/3} ~\left(\frac{6 \pi
p}{k_n}\right)^{1/3}. \label{emt_mu}
\end{equation}

{(2) \it  Path-dependent models, $k_t\neq0$, frictional grains}: In this case
tangential elastic forces are taken into consideration. In
principle the energy functional (\ref{energy}) now depends on the
path. However, it has been shown that the second order elastic
constants are still path-independent under the framework of EMT,
while path-dependency appears only in the third order elastic
constants \cite{johnson1}.

The bulk modulus is not affected by the introduction of tangential
forces, and Eq. (\ref{emt_k}) is still valid in this case.
However, the shear modulus is modified  according to:
\begin{equation}
\mu_e(p)= \frac{k_n + \frac{3}{2} k_t}{20 \pi}~  (\phi Z)^{2/3}
~\left(\frac{6 \pi p}{k_n}\right)^{1/3}. \label{emt_mu2}
\end{equation}

The above results have been obtained by a number of authors using
different methods and are valid in 3-D
\cite{digby,walton,johnson1}. It should be noted that the above
results are obtained for a system of infinitely rough spheres,
i.e. when $\mu_f\to \infty$. Thus, there is no sliding, and
Coulomb friction is not considered, although there is a tangential
elastic restoring force as given by Eq. (\ref{ft}). See Appendix A
for further discussion.

The $p^{1/3}$ dependence in Eqs. (\ref{emt_k})-(\ref{emt_mu2}) is
a direct consequence of the scaling of the normal Hertz force on
the deformation. Since
\begin{equation}
p \sim F_n \sim \xi^{3/2} \sim \epsilon^{3/2}.
\end{equation}
Then we expect
\begin{equation}
\mu_e \sim K_e \sim \frac{\partial p }{\partial \epsilon} \sim
\epsilon^{1/2} \sim p^{1/3}.
\end{equation}

We note that a system of linear springs, $F_n\sim \xi$, would give
rise to elastic constants which are independent of pressure, as in
the linear elasticity theory.

\subsection{ Discrepancies between theory and experiments.}

It was found experimentally that the shear and bulk moduli of an
assembly of spherical grains vary with the confining stress, $p$,
faster than the $p^{1/3}$ power law predicted by Eqs.
(\ref{emt_k})-(\ref{emt_mu2})
 \cite{goddard}.
Another way of seeing the breakdown of the elastic theory is to
focus on the ratio $K / \mu$.
According to Eqs. (\ref{emt_k}) and (\ref{emt_mu2}) for frictional
grains
\begin{equation}
\frac{K_e}{ \mu_e} =\frac{5 (2-\nu_g)}{3 (5-4\nu_g)},
\label{kovermu}
\end{equation}
 independent of stress, a value which
depends only on the Poisson ratio of the bead material.  The
experiments give $K/\mu\approx 1.1-1.3$ \cite{domenico,yin}.
EMT predicts $K_e/\mu_e=0.71$, if we take $\nu_g=0.2$ for the
Poisson ratio of glass. [The EMT prediction is rather insensitive
to variations of $\nu_g$; $K_e/\mu_e = 0.71 \pm 0.04$ for $\nu_g =
0.2 \pm 0.1$.] Conversely, a value $\nu_g\simeq 1.2$ would be
needed in order to fit the experimental  $K/\mu$, clearly
violating the upper thermodynamic limit of $\nu_g\le 1/2$
\cite{landau}.

Another quantity of interest is the effective Poisson ratio of the
pack $\nu$. According to EMT, $\nu_e$ is again independent of
pressure and given only in terms of the Poisson's ratio of the
grains:
\begin{equation}
\nu_e \stackrel{def}{=} \frac{K_e - 2/3 \mu_e}{2 ( K_e - 1/3
\mu_e)} = \frac{\nu_g}{2(5-3\nu_g)}. \label{nu_e}
\end{equation}

Thus, for typical glass beads ($\nu_g=0.2$) we find a predicted
value  of $\nu_e = 0.02$ which is one order of magnitude smaller
than typical experimental values $\nu \approx 0.28$
\cite{domenico}; another  serious disagreement.

The origin of the above discrepancies has not been clear: it could
be due to the breakdown of the Hertz-Mindlin force law at each
grain contact, or it could be associated with the breakdown of the
elasticity theory applied to granular systems. De Gennes
\cite{pgg} proposed that a thin shell of oxide layer would give a
faster growth with stress of the elastic moduli of the system,
which may explain the behavior for metallic beads.  Goddard
\cite{goddard} proposed that sharp angularities of the grains (for
instance sand grains) may modify the contact force law between
grains, giving rise to a different stress dependence.  Other
authors \cite{goddard,jenkins} have suggested that the increasing
number of contacts with stress may be the reason for the
discrepancies in the stress dependence of the moduli.  Jenkins
{\it et al.} \cite{jenkins} measured the elastic moduli using
numerical simulations for a single pressure and concluded that EMT
does not correctly describe the shear modulus but it describes the
bulk modulus fairly well. Other experimental work done by Liu and
Nagel \cite{liu1} and Jia {\it et al.} \cite{jussieu} concentrated
on the role played by force chains in sound propagation. Different
approaches for one dimensional elastic chains have also been
applied to wave propagation in granular media
\cite{nesterenko,hascoet}.

\subsection{Linear viscoelastic constitutive models}
\label{viscoelastic}



In order to understand our results it is important to generalize
the elastic concepts introduced above to the full viscoelastic
response.
In linear viscoelasticity \cite{ferry}, the current state of
stress specified by the stress tensor $\sigma_{ij}$ is determined
by the past history via a linear constitutive equation
\begin{equation}
\sigma_{ij}(t) = \int_{-\infty}^t G_{ijkl}(t-t')~
\dot\epsilon_{kl}(t')~ dt', \label{const}
\end{equation}
where  $\dot\epsilon_{kl}=\partial\epsilon_{kl}/\partial t$ is the
strain rate,  and $G_{ijkl}(t)$ is called the relaxation modulus
tensor.

For an isotropic linear viscoelastic material, the relaxation
modulus tensor has only two independent components. These are the
shear relaxation modulus $G(t)$ and the bulk relaxation modulus
$K(t)$ characterizing the response to shear $\dot\epsilon_{12}$
and bulk deformation  $\dot\epsilon_{ii}$.
The relaxation modulus $G(t)$ and $K(t)$ are conceptualized as the
time-dependent analogues of the shear $\mu_e$ and bulk modulus
$K_e$ in elasticity theory.

In this study we will concentrate on the stress relaxation after a
sudden strain imposed via a simple shear, a pure shear, or a
uniaxial compression.  For example, a shear strain is applied
instantaneously, at $t=0$, from its initial value of zero to a
final, constant value, $\epsilon_{12}$. For this situation we have
$\dot\epsilon_{12}(t) = \epsilon_{12} \delta (t)$.  Equation
(\ref{const}) reduces to 
\begin{equation}\sigma_{12}(t) =
\epsilon_{12} G(t)\:\:\:.\label{fred}\end{equation} 
Therefore,
this strain protocol immediately yields complete information on
the response function, $G(t)$, [shorthand for $G_{1212}(t)$ here]
simply by measuring $\sigma_{12}(t)$. This is a strain protocol
which is particularly simple to implement in our MD simulations.

For a perfectly elastic solid  the relaxation modulus is
independent of time, $G(t) = G = $constant, and one can define the
shear modulus of the solid as
\begin{equation}
\mu_e \stackrel{def}{=} \sigma_{12}/\epsilon_{12}=G.
\end{equation}
We will show that the instantaneous response of the viscoelastic
granular material, $G(t=0)$, represents the shear modulus,
$\mu_e$, as calculated by the effective medium theories of
continuum elasticity.

For a Newtonian liquid $G(t) = \eta \delta(t)$, where $\eta$ is
the viscosity.  For a viscoelastic liquid, $G(t)$ approaches zero
as $t \to \infty$.  For a viscoelastic solid, structural
relaxation and elasticity lead to a finite modulus as $t \to
\infty$:
\begin{equation}
\mu = G(t \to \infty).
\label{mu1}
\end{equation}
A similar analysis can be performed for the bulk modulus, $K(t)$
defined as
\begin{equation}\sigma_{ii}(t) = 3
\epsilon_{ii} K(t)\:\:\:.\label{fred2}\end{equation} 
to obtain $K(0) = K_e$ and 

\begin{equation}
K = K(t\to\infty).
\label{k1}
\end{equation}

Equations (\ref{mu1}) and (\ref{k1}) will be used to calculate
the moduli in the simulations.


\subsection{Molecular Dynamics simulations}
\label{md}

In MD simulations of granular matter the net force and moment on
each grain depend on the choice of intergrain contact laws
\cite{cundall,wolf-md}. Here, we follow the Discrete Element
Method (DEM) developed by Cundall and Strack \cite{cundall} and
solve Newton's equations for an assembly composed of soft
elasto-frictional spheres interacting via Hertz-Mindlin contact
forces and Coulomb friction as described in Section
\ref{background-forces} \cite{johnson}. We employ a
time-stepping, finite-difference approach to solve the Newtonian
equations of motion simultaneously for every grain in the system:
\begin{equation}
\mbox{\bf F}=m ~ \ddot{\mbox{\bf x}}, \label{RFDF}
\end{equation}
\begin{equation}
\mbox{\bf M}=I ~ \ddot { \mbox{\boldmath{$\theta$}}}, \label{RFDM}
\end{equation}
where {\bf F} and {\bf M} are the net force and moment acting on a
given grain, $m$ and $I$ are the mass and moment of inertia, and
$\ddot{\mbox{\bf x}}$ and $\ddot { \mbox{\boldmath{$\theta$}}}$
are the linear and angular accelerations of the grain,
respectively.

The numerical solution of Eqs. (\ref{RFDF}) and (\ref{RFDM}) are
obtained by integration, assuming constant velocities and
accelerations for a given time step: linear and angular velocities
are determined from the knowledge of the force and torque, and
grain displacements and rotations at the next time step are
calculated from the average velocities. Grain motions can be
initiated by gravitational forces, by external forces prescribed
by stress or strain rate boundary conditions, and by forces
resolved at intergrain contacts. Strain rates are assumed to be
low, and small time steps $\Delta t$ are chosen to ensure that the
disturbance of a given grain only propagates to its immediate
neighbors (see Appendix D).

{\it Viscous damping.---} Damping of grain motions must be
included in the calculations to prevent the continuous oscillation
of an elastic system.  Although damping is a physical reality, and
physically meaningful mechanisms might well be incorporated, our
concern here is to get the simulations to equilibrate to the final
answer in a reasonable amount of computer time.

Several damping methods are possible. Global damping considers the
particles immersed in a viscous fluid and is provided by
introducing viscous force terms in Eqs.  (\ref{RFDF}) and
(\ref{RFDM}). These drag forces are proportional to the absolute
velocity and angular velocity of the particles: $\sim - \gamma_n
\dot{\mbox{\bf x}}$, and $\sim -\gamma_t \dot{
\mbox{\boldmath{$\theta$}}}$, where the $\gamma$'s are the global
damping coefficient related to the viscosity of the immersing
fluid (which could be for instance air).

Global damping is introduced to guarantee that the system can
reach an equilibrium state with zero velocity at a given pressure.
Its physical significance is being studies at the moment
by experiments and computer simulations.
Another source of damping implies a contact force term acting at
every contact point, proportional to the relative velocities of
the grains. Microscopic contact damping occurs due to the viscous
dissipation of energy in the bulk of the particle material when
they are deformed and it may also occur if 
liquid bridges are formed at the contact points between the particles.
Here, a damping force is added to each contact
force, Eqs. (\ref{fn}) and (\ref{Ft}), proportional to the
relative normal and shear velocities, $\beta_n \dot \xi$ and
$\beta_t \dot{s}$, respectively, with  $\beta_n$ and $\beta_t$ the
contact damping coefficients.
Typical values of the damping constant are given in
\cite{wolf-md}.

In this study we will use global damping for the preparation
of the sample and the  calculation of the
elastic constants. This procedure is necessary to achieve the
final equilibrium states which we wish to explore 
(see Appendix B for a discussion).


{\it Computation of stress.---} The macroscopic stress tensor for
point contacts in a volume $V$ is given by
\cite{digby,walton,johnson1}
\begin{equation}
\sigma_{ij}=\frac{1}{2 V}\sum_{contacts}(F_{i} R n_j + R n_i F_j),
\label{stress}
\end{equation}
where $\hat{\bf n}$ is the unit vector joining the center of two
spheres in contact.

\section{Acoustic Experiments}
\label{experiments}

In the simplest experiments, a packing of glass beads is confined
under hydrostatic conditions and the compressional and shear sound
speeds, $v_p$ and $v_s$, are measured as functions of $p$
\cite{goddard,domenico,yin,winkler}.

In the long-wavelength limit, the sound speeds are related to the
elastic constants of the aggregate by Eqs. (\ref{vp}) and
(\ref{vs}). Here we perform our own experiments according to
standard sound propagation techniques \cite{domenico,winkler}.


\subsection{Experimental Configuration}

We used a set of high quality glass beads of a small enough
diameter to measure an appreciable signal at low pressure.  From
the experimental data of Domenico \cite{domenico}, we expect
compressional velocities  $v_p \sim 1000$ m/s and
shear velocities $v_s \sim 500$ m/s at low pressures.  We perform
ultrasonic measurements with pulses of frequency $f=500$ kHz, and
we find that the maximum size of the beads should be $R \ll v_s /
(2 f)$. Then, we choose a set of glass beads of diameter 45
$\mu$m in order to reach the desired low pressures.

\begin{figure}
\centering {\resizebox{9cm}{!}{\includegraphics{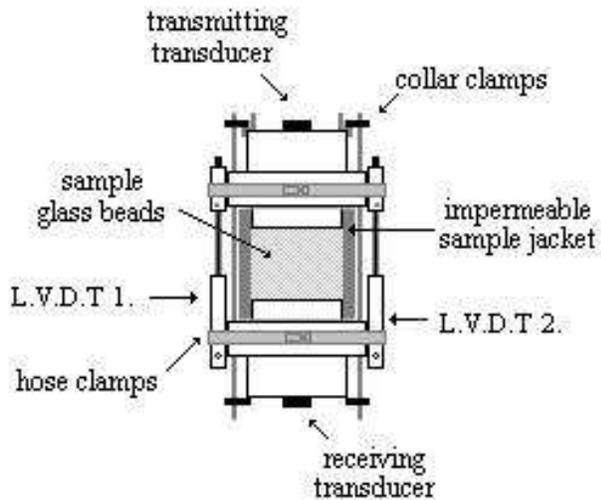}}}
\caption{Container and Transducers-LVDT apparatus used in the sound
propagation experiments.} 
\label{setup}
\end{figure}

The glass beads were cleaned and dried to avoid any agglomeration
(electrostatic forces or moisture). The glass beads were then
deposited into a flexible container (Tygon sleeve) of 3 cm height
and 2.5 cm radius. Transducers and a pair of linear variable
differential transformers (LVDT, for measurement of displacement)
were placed at the top and bottom of the flexible membrane (see
Fig. \ref{setup}).

Before starting the measurements, a series of tapping and
vibrations were applied to the container in order to let the
grains settle into the densest possible packing.  Our goal is to
establish the sample in the reversible state described by, e.g.
Figure 2 of Nowak, {\it et al} \cite{nowak}.  The entire system
was then put into a pressure vessel filled with oil (see Fig.
\ref{setup}). We then applied confining pressures ranging from 0
to 140 MPa. The pressure was cyclically applied several times
until the system exhibited minimal hysteresis.  At this point
shear and compressional waves were propagated by applying pulses.
The sound speeds and corresponding moduli were obtained by
measuring the arrival time from ``head to head" of the transducers
for the two sound wave types.


\subsection{Acoustic measurements}

\begin{figure}
\centering {\resizebox{8.5cm}{!}{\includegraphics[angle=-90]{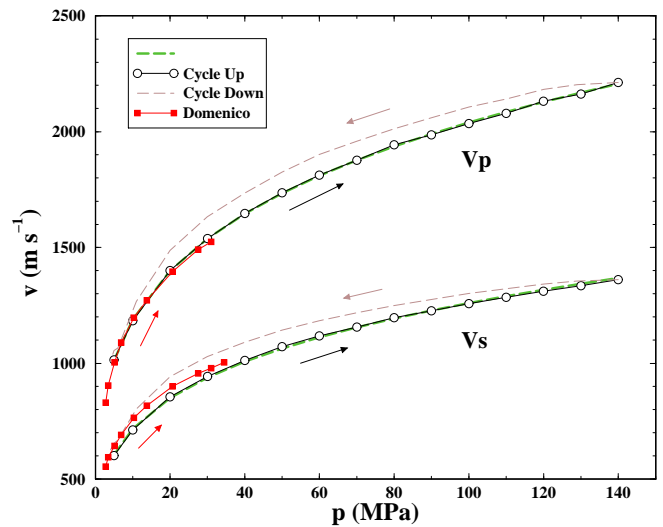}}}
\caption{(a) Wave velocities versus pressure obtained in our experiments.
Also shown are the results of Domenico \protect\cite{domenico} 
for comparison. We cycle up and down in pressure to avoid hysteresis.}
\label{velocities_linear}
\end{figure}

The results we obtain are plotted in Fig. \ref{velocities_linear}
and they compare well with the available data of Domenico
\cite{domenico} for the range  0-40 MPa. There remains an
hysteresis component between the cycle upwards and downwards in
pressure which is representative of the packed system. A more
detailed comparison with theory and simulations is done in Figs.
(\ref{exp}) and (\ref{mu_1/3}), below.

Because of the deformation of the glass beads the height of the
system decreases with the increasing pressure. In order to obtain
the correct velocities from the arrival time of the signal, we
accurately measure the displacement of the transducers as the
pressure is increased with a pair of LVDTs. In order to avoid
fracture of the particles due to the external pressure we use
small particle sizes to reduce the intensity of the contact
forces. To get a qualitative idea of the presence of crushing
within the system we observe the sample under a microscope after
the experiments. We find that crushing occurs only in a very small
fraction of the beads. When the experiment was repeated with
larger beads of diameter 0.3 mm many particles appeared to be
crushed after applying pressures of 140 MPa. Moreover, during this
test there was a strong acoustic emission and a severe inflection
in the sound speeds could be noticed as we increased the pressure
during the first cycle upwards due to the crushing of the beads.
For the beads of size $45 \mu$m, no inflection is observed and no
acoustic emission is heard during the experiment.


\section{Numerical simulations}
\label{numerics}

We perform MD simulations of a system of 10000 spherical particles
in a periodically repeated cubic cell of approximately 4mm sides.
The particles interact via Hertz-Mindlin contact forces and we
choose typical values for glass beads for $\mu_g=29$ GPa and
$\nu_g = 0.2$ for a close comparison with experiments.  We assume
a distribution of grain radii in which $R_1=0.105$ mm for half the
grains and $R_2=0.095$ mm for the other half.  Our results are
quite insensitive to the choice of the size distribution. We
include viscous damping terms to allow the system to relax toward
static equilibrium as discussed in Section \ref{md}.

The general scheme of the simulations is as follows: The
simulations begin with a gas of 10000 grains distributed at random
positions inside the cubic cell. We first apply a compression
protocol so that a dense random packing is generated corresponding
to a predetermined value of the pressure. Then, an incremental
infinitesimal compression or shear is applied to the unit cell and
the change in stress is computed, once the system re-equilibrates.
Thus, we obtain the bulk and shear moduli for the system at each
confining pressure.

\subsection{The reference state: numerical protocol}

One of the critical issues in this study is how to obtain a proper
rigid frame of reference Eq. (\ref{R}), {\bf \{R\}}, from where we
could calculate the elastic moduli. Our calculations begin with a
numerical protocol designed to mimic the experimental procedure
used to prepare dense packed granular materials at a given
confining pressure. In the experiments, the initial bead pack is
subjected to mechanical tapping and ultrasonic vibration in order
to increase the solid phase volume fraction, as discussed in the
previous section.

During the numerical preparation stages we turn off the transverse
force between the grains ($k_t = 0$); because there are no
transverse forces, the grains slip without resistance and the
system reaches the high volume fractions found experimentally
during the initial compression process.
 We found that by preparing the system with frictional
and elastic tangential forces, the system reaches states of lower
volume fraction. A more complete study of this effect will be
presented in Part II \cite{unknownauthors}. In the following
calculation we concentrate on the preparation without friction, so
that we can obtain the most compact states possible, mimicking our
experimental procedure. We then restore the tangential Mindlin
force and friction when we calculate the elastic constants.

\begin{figure}
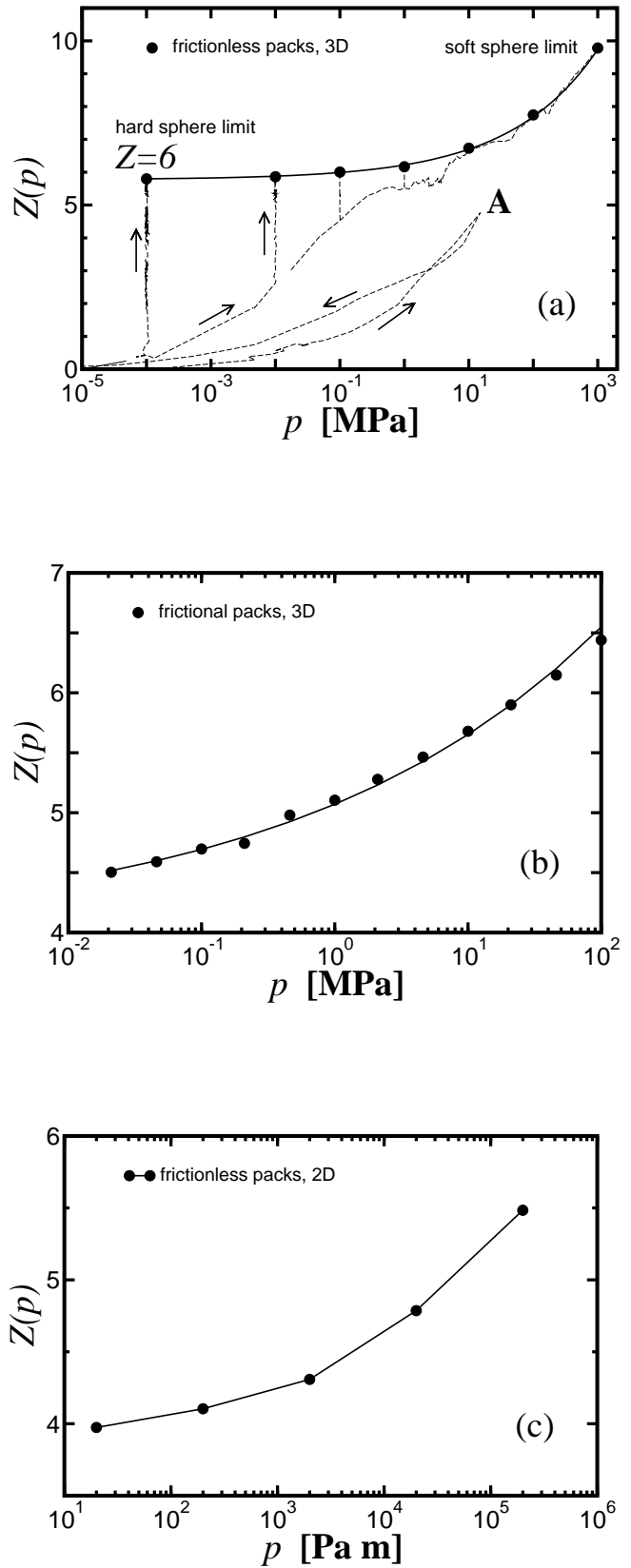

\centering
{\resizebox{8.5cm}{!}{\includegraphics{fig3a.eps}}
}
\vspace{1cm}

\centering
{\resizebox{8.5cm}{!}{\includegraphics{fig3b.eps}}
}
\vspace{1cm}

\centering
{\resizebox{8.5cm}{!}{\includegraphics{fig3c.eps}}
}
\caption{Coordination number versus pressure obtained in the simulations.
 (a) Frictionless
packs in 3D. The system becomes isostatic as $p\to 0$ and 
$Z\simeq 6$. (b) Frictional
packs in 3D. Is the isostatic limit $Z_c=4$ reached asymptotically
as $p\to 0$? See Part II \protect\cite{unknownauthors} for details. (c)
Frictionless packs in 2D. Here the system is isostatic with
$Z_c \simeq 4$ as $p \to 0$.} 
\label{coord}
\end{figure}

Starting with a set of non-contacting particles, we first apply a slow
compression to bring the particles closer until a specified value
of the pressure and coordination number is attained. This initial
compression is specified by the dashed lines in Fig. \ref{coord}a.
If the compression is stopped just before reaching a volume
fraction of random close packing (specified as Point A in Fig. \ref{coord}a)
and the system is allowed
to relax, then system will relax to zero pressure
and zero coordination number, since it cannot equilibrate below
the maximum close packing fraction. This is indicated in Fig.
\ref{coord}a as the decrease of the coordination number and
pressure towards zero. The
compression is then continued to a point above the critical
packing fraction at a target pressure, $p_t$. The target pressure,
is maintained with a ``servo'' mechanism \cite{cundall}
which constantly adjusts the applied strain rate  $\dot \epsilon$
until the system reaches equilibrium at $p_t$ according to the
following prescription:
\begin{equation}
\dot \epsilon = g (p - p_t),
\end{equation}
where $p$ is the actual pressure of the system and $g$ is a gain
factor which is tuned to achieve equilibrium at every given
pressure in an optimal way.


\subsection{Coordination number}

The above protocol is repeated for different target pressures and we
obtain the average coordination number $Z$ of these equilibrium
states as a function of the pressure, as seen in Fig. \ref{coord}a.
Several important points can be seen from this plot. Firstly, the
average coordination number increases with the pressure as
expected. Secondly, we find that the coordination number of the
pack approaches a critical minimal value close to $Z_c\approx 6$
as $p \to 0$. At low pressures, compared to the shear modulus of
the beads ($p\ll 26$GPa), the system behaves more like a pack of
rigid balls. At this point the beads are minimally connected at
 $Z_c\approx 6$, while in two dimensions (see Appendix E) 
the same preparation 
protocol gives
$Z_c\approx 4$ (Fig. \ref{coord}c).

Such low coordination numbers can be understood in terms of simple
constraint arguments for a system of $N$ frictionless rigid
particles in $D$ dimensions
\cite{alexander,isostatic,edwards-grinev,ball}. We need to
determine $Z N/2$ normal forces with $D N$ equations of force
balance. We find a critical coordination number for which the
equations of force balance are soluble 
as $Z_c=2 D$. For large values of the
confining pressure more grains are brought into contact, and the
coordination number increases from its minimal value required for
stability; the system is underconstrained. Empirically, we find in 3D

\begin{equation}
Z(p) = Z_c + \left(\frac{p}{ ~\mbox{10 MPa}}\right)^{0.30(5)}.
\label{Z}
\end{equation}

The pressure 10 MPa is significant since it
determines the characteristic pressure of the 
crossover from the minimal coordination number to a
larger one.

We also measure the volume fraction as a function of pressure and
find that it approaches a critical value of $\phi_c\approx 0.63$
in the rigid ball limit as $p \to 0$:
\begin{equation}
\phi(p) = \phi_c + \left(\frac{p}{ ~\mbox{14 GPa}}\right)^{0.62(6)}.
\label{phi-p}
\end{equation}
The value of $\phi_c=0.63\approx \phi_{\mbox{\scriptsize RCP}}$
corresponds to the volume fraction at random close packing (RCP):
the densest possible random packing of hard spheres
\cite{bernal,finney,berryman}, since the hard sphere limit in our
system of deformable particles is achieved when the pressure
(deformation) vanishes (RCP is only achieved asymptotically in our
simulations). The exponent 0.62 is consistent with dimensional
arguments which would predict a value inverse of the 
power law between the force and displacement in the Hertz law, 
i.e. a 2/3  exponent.
The exponent in Eq. (\ref{Z}) is determined by the behavior of the
pair distribution function near jamming.
These exponents agree with similar calculations done by
O'Hern {\it et al.} \cite{ohern},
 and they will de discussed
in more detail in Ref. \cite{unknownauthors}.

The low value of $Z_c$ is very significant (this number should be
compared, for instance, to $Z=12$ for a FCC packing) because
at this minimal coordination the equations for the
force distribution can be solved without reference to the state
of strain in the system. This is the isostatic limit
\cite{alexander,isostatic} and the starting point of recent
theories of stress distributions in granular packs
\cite{edwards-grinev,fragile,ball}. Concepts such as
fragility and marginal rigidity depend on the existence of this
minimally connected state. In the conclusions we will come back to
discuss this issue. As previously reported in \cite{mgjs} and
\cite{makse-fc}, Eq. (\ref{Z}) provides a numerical evidence of the
existence of the minimally connected state in frictionless
granular packs. For other numerical work see \cite{silbert}.

To test the robustness of these results, we have employed a second
protocol in which the system is prepared by compressing to a point
beyond the RCP fraction, then letting the grains relax to
equilibrium without the servo mechanism.  The final $Z(p)$ curve
is essentially identical to the one shown in Fig. \ref{coord}a. For
this reason we believe that we have accurately approximated the
reversible state of dense random packing, in the sense discussed
by Nowak, {\it et al.} \cite{nowak}.

It is important to recall that the above results have been
obtained for a system without friction. A similar preparation
protocol for grains with friction gives rise to different packings
with lower coordination number. Similar constraints arguments as
explained above give $Z_c=D+1$ for this case. Fig. \ref{coord}b
shows $Z(p)$ obtained for a system with friction showing that a
minimal $Z_c\approx 4$ in 3D may be approached asymptotically as $p\to
0$, although at a slower rate than in the frictionless case. Is
the $Z_c=4$ isostatic limit achieved as $p\to 0$? We have given a
positive answer to this question in \cite{makse-fc}. However,
recent studies \cite{silbert} suggest that this may not be the
case. We refer the interested reader to Part II of this work
\cite{unknownauthors} for our more recent results
showing that the rate of compression (analogous to the rate of cooling of a
a glass-forming
liquid below the glass transition) plays
a significance role in achieving the isostatic limit in frictional packs.
 From now on we
will concentrate on the calculation of the elastic properties of
granular media using the states depicted in Fig. \ref{coord}a as
our starting point. Of course, we will restore $k_t \neq 0$ for
the calculation of the moduli.

\subsection{Calculation of elastic moduli with MD}

Consider the calculation of the elastic moduli of the system as a
function of pressure.  Beginning with the equilibrium states of
Fig. \ref{coord}a we first restore the transverse component of the
contact force  by setting $k_t \ne 0$. We then apply a small
perturbation to the system and measure the resulting response. We
don't expect slippage to occur since we apply infinitesimal strain
perturbations, but since we deal with a finite system we set the
friction coefficient $\mu_f$ to a large value to avoid sliding at
the contacts. The elastic moduli are calculated by applying a
given affine infinitesimal strain perturbation $\Delta \epsilon$
as given by Eq. (\ref{affine}) and then monitoring the response of
the corresponding stress $\sigma(t)$ as a function of time. After
the system equilibrates again as $t \to \infty$, the moduli are
obtained from Eqs. (\ref{mu1}) and (\ref{k1}) as the 
change in stress between the final state and the stress before the
perturbation  $\Delta \sigma / \Delta \epsilon$. The procedure is
repeated for $\Delta \epsilon \to 0$ to guarantee that we are
testing the linear response regime where the elastic moduli become
independent of $\Delta \epsilon$. Interestingly we find that the
region where the elastic constants are well defined decreases as
the pressure decreases. This is in agreement with the prediction
of the EMT for the 3rd order elastic constants which are found to
diverge as $\sim \epsilon^{-1/2} \sim p^{-1/3}$ \cite{johnson1}.

The shear modulus is calculated from a simple shear test ($\Delta
\epsilon_{12} = \Delta \epsilon_{21} \neq 0$) as given by Eq. 
(\ref{stress-strain})

\begin{equation}
\mu=\frac{1}{2}\frac{\Delta\sigma_{12}}{\Delta\epsilon_{12}},
\label{NPST}
\end{equation}
and also from a pure shear test with $\Delta \epsilon_{11} = -
\Delta \epsilon_{22}$:

\begin{equation}
\mu=\frac{1}{2}\frac{(\Delta\sigma_{22}-\Delta\sigma_{11})}{(\Delta
\epsilon_{22}-\Delta\epsilon_{11})}. \label{NBT}
\end{equation}
We find that the values of $\mu$ determined from these two methods
agree with each other, as expected for an isotropic system.

The bulk modulus is obtained from a uniaxial compression test
along the 1-direction and keeping the strain constant in the other
directions $\Delta \epsilon_{22} = \Delta \epsilon_{33} = 0$, and
$\Delta \epsilon_{11}\neq 0$:

\begin{equation}
K=\frac{\Delta\sigma_{11}}{\Delta \epsilon_{11}}-
\frac{4}{3}\mu\:\:\:. \label{UCT}
\end{equation}

Here the stress, $\sigma_{ij}$, is determined from the measured
forces on the grains Eq. (\ref{stress}), and the strain,
$\epsilon_{ij}$, is determined from the imposed dimensions of the
unit cell. For instance  $\epsilon_{11} = \Delta L /L_0$ where
$\Delta L$ is the infinitesimal change in the $11$ direction and 
$L_0$ is the size of the reference state at the given $p$.

The results of our numerical calculations for
$K(p)$ and $\mu(p)$ are shown in Fig. \ref{exp}.
These results have been obtained for packings of 10000
particles. Calculations done with 432 spheres show similar values
indicating that the results are free of finite size effects.
We see that our experimental and numerical
results are in reasonably good agreement. Also shown are data
measured by Domenico ~\cite{domenico}. Clearly, the experimental
data are somewhat scattered at low pressure. It reflects the
difficulty of the measurements, especially at the lowest pressures
where there is a significant signal loss.  Nevertheless, our
calculated results pass through the collection of available data.
It should be noted that the experiments are compared against the
numerical results without resorting to the use of fitting
parameters, since all the constants characterizing the grain
material ($\mu_g$ and $\nu_g$) are known from the properties of the grains.


\subsection{Breakdown of the EMT. Problems with  $\mu$}

\begin{figure}
\centering {\resizebox{8.5cm}{!}{\includegraphics{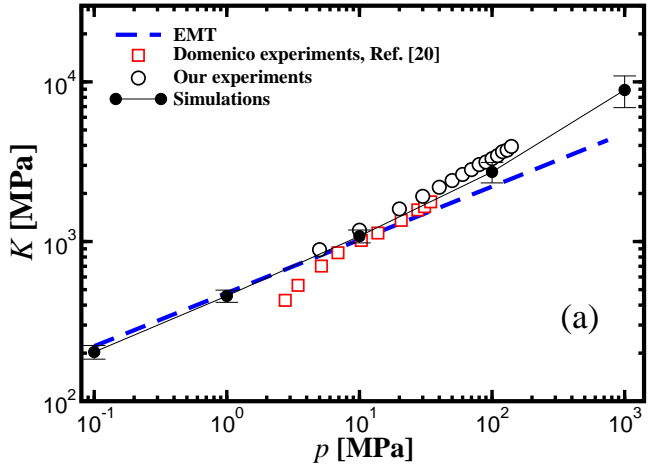}}}
\vspace{2cm}

\centering {\resizebox{8.5cm}{!}{\includegraphics{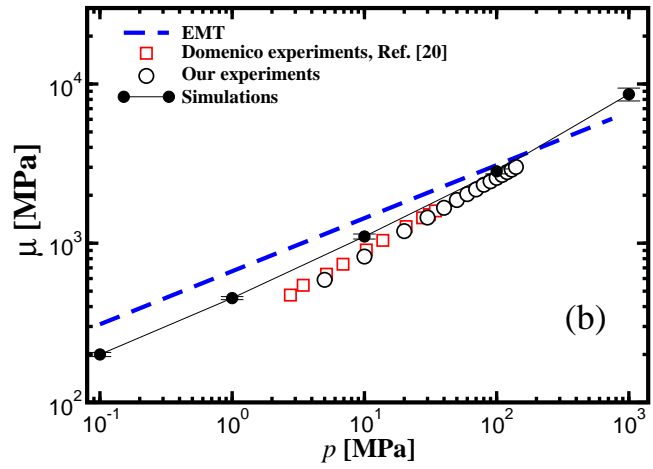}}}
\caption{Pressure dependence of the elastic moduli, (a) bulk and
(b) shear moduli from MD, our experiments, Domenico experiments
and EMT.} \label{exp}
\end{figure}

Also shown in Fig. \ref{exp} are the EMT predictions Eq.
(\ref{emt_k}) and (\ref{emt_mu2})
 using the same parameters as in the simulations.
We set $Z=6$ and $\phi=0.64$, independent of pressure. At low
pressures we see that $K$ is well described by EMT.  At larger
pressures, however, the experimental and numerical values of $K$
grow faster than the $p^{1/3}$ law.  The situation with the shear
modulus is even less satisfactory.  EMT overestimates $\mu (p)$ at
low pressures but, again, underestimates the increase in $\mu (p)$
with pressure.

\begin{figure}
\centering {\resizebox{8.5cm}{!}{\includegraphics{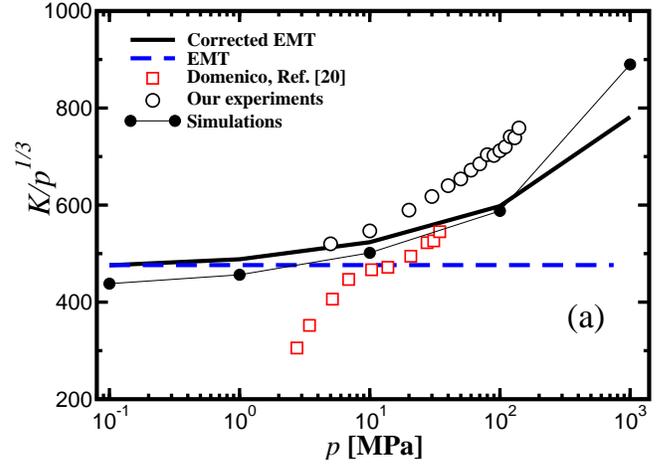}}}
\vspace{1cm}

\centering {\resizebox{8.5cm}{!}{\includegraphics{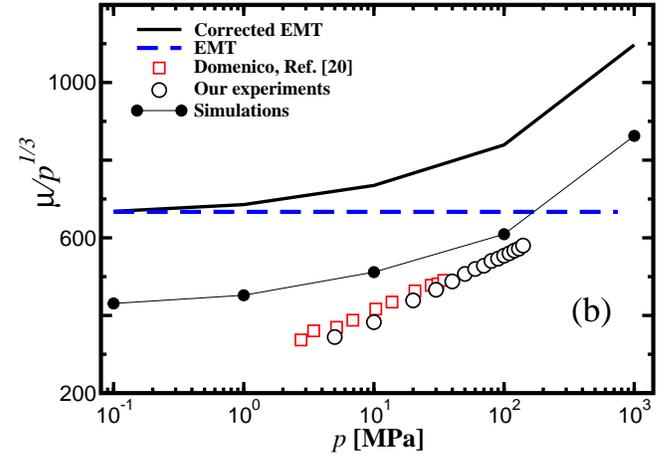}}}
\caption{Elastic moduli, (a) bulk and (b) shear, normalized to
$p^{1/3}$ and corrected EMT taking into account the pressure
dependence of $Z(p)$ from Fig. \ref{coord}a as well as
$\phi(p)$.} \label{mu_1/3}
\end{figure}

To investigate the failure of EMT in predicting the correct
pressure dependence of the moduli, we re-plot the  moduli divided
by $p^{1/3}$ in Fig. \ref{mu_1/3}.

For such a plot, EMT predicts a horizontal straight line but we
see that the numerical and experimental results are clearly
increasing with $p$. It is tempting to try to fit the data with
another power law. However, we must first include the power law
dependence of the coordination number and the volume fraction 
with the pressure as
given by Eqs. (\ref{Z}) and  (\ref{phi-p}). Thus we modify Eqs.
(\ref{emt_k}) and (\ref{emt_mu2}) to include the pressure
dependence $Z(p)$ and  $\phi(p)$ (this latter is a much smaller effect, 
see below).
The corrected EMT is also plotted in Fig. \ref{mu_1/3} and we see
that it predicts the same trend with pressure as the simulations.
The experimental data also seem to be following this trend but
more data over a larger pressure range are clearly needed.

When analyzing $K(p)$, we find that the corrected EMT is in
essentially exact agreement with our numerical simulations and
experimental data. Thus, we tend to conclude that the anomalous
scaling found in the experiments is be a measurement of a
crossover behavior as obtained by combining Eqs. (\ref{emt_k}) and
(\ref{emt_mu2}) with the nonlinearity of Eqs. (\ref{Z})
 giving rise to two distinct scaling regimes:

\begin{eqnarray}
\label{scaling}
K(p) & \sim & \mu(p)  \sim  p^{1/3}, ~\mbox{for $p \ll 10$ MPa},
\\
K(p) & \sim & \mu(p)  \sim p^{5/9}, ~ \mbox{for 10 MPa $ \ll p \ll 14$ GPa}.
\nonumber
\end{eqnarray}

Here we have not included the pressure dependence
of the volume fraction Eq. (\ref{phi-p}) since it appears at the 
very large pressures
above 14 GPa. At these pressures the beads are not supposed to follow anymore
the Hertz law (and they may, in fact, fracture). Therefore we exclude 
this regime from
our scaling analysis in Eq. (\ref{scaling}).

Since the experiments are usually done near the crossover pressure
of 10 MPa, it holds to reason that they could be  measuring
 a crossover behavior rather
than a true scaling regime. Moreover, even for 
pressures larger than 10 MPa the Hertz contact 
mechanics approach might fail 
since
the Hertz theory is based on small perturbations. Thus, the true
final scaling regime Eq. (\ref{scaling}) might not be accessible
experimentally, at least for glass beads and other rigid
materials. It would be interesting to see if such a crossover
could be observed in softer materials.

The substitution of Eqs. \rf{Z} and \rf{phi-p} into \rf{emt_k} and
\rf{emt_mu2} is something of an ad hoc procedure; Eqs. \rf{emt_k}
and \rf{emt_mu2} were derived under the assumption that $Z$ and
$\phi$ are stress-independent quantities.  Within the context of
the affine assumption, the EMT derivation can be modified to
account for a continuously changing coordination number, $Z(p)$.
Let us assume that, in the limit of zero pressure, there is a
probability distribution $P(h)$ of gap sizes, $h$, between each
ball and its neighbors:

\begin{equation} P(h) = Z_c\delta(h) + a_1 + a_2 h + ... \label{gap}
\end{equation} where $Z_c = 6$ represents the coordination number at zero
stress and the rest is a Taylor's series expansion around $h=0$.
It is straightforward to re-do the derivations leading to Eqs.
\rf{emt_k} and \rf{emt_mu2} following the prescription in, e.g.
\cite{johnson1}.  The results, expressed in terms of the static
compressive strain, $\epsilon<0$, are \begin{equation} p =
\frac{\phi k_n}{6\pi} \left [ Z_c (-\epsilon)^{3/2} +
\frac{2}{5}(a_1 R)(-\epsilon)^{5/2} + ... \right] \label{p_gap}
\end{equation} \begin{equation} K = \frac{\phi k_n}{12\pi} \left [ Z_c
(-\epsilon)^{1/2} + \frac{2}{3}(a_1 R)(-\epsilon)^{3/2} + ...
\right] \label{K_gap} \end{equation} Using a judiciously chosen
value of $a_1 \neq 0$, and neglecting $a_2$ and all higher-order
terms, a cross-plot of Eq. \rf{K_gap} against \rf{p_gap} mimics
the molecular dynamic simulations in Figure \ref{exp}.  We note,
however, that, taken literally, Eq. \rf{gap} predicts $Z-Z_c
\propto p^{2/3}$ for small $p$, in contrast to Eq. \rf{Z}.

Since the bulk modulus is approximately described by the corrected
EMT, throughout the rest of the paper we focus on $\mu(p)$. In
Fig. \ref{mu_1/3} it is shown that even though the pressure trend
is well described by the corrected EMT, the theory still
overestimates the value of the shear modulus. We will see later
that the overestimation depicted in Fig. \ref{mu_1/3} becomes
enormous when the tangential forces are diminished towards zero.
In this limit, the breakdown of the EMT is clearly established.

\begin{figure}
\centering {\resizebox{8.5cm}{!}{\includegraphics{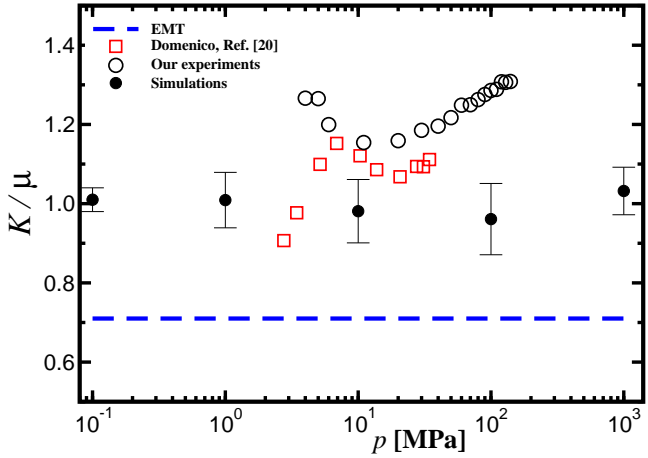}}}
\caption{Ratio $K/\mu$ for MD, experiments, and EMT (pressure
independent)} \label{kovermu_fig}
\end{figure}

Another way of seeing the breakdown of EMT is to focus on the
ratio $K/\mu$, which is independent of pressure in the theory Eq.
(\ref{kovermu}), the simulations, and approximately so in the
experiments, as seen in Fig. \ref{kovermu_fig}. [The variation at
low pressure may reflect the difficulty in propagating sound at
low confining pressures]. The experiments give $K/\mu\approx
1.1-1.3$. Our simulations give $K/\mu\approx 1.05\pm0.05$ in good
agreement with experiments.  Notice, however, that the EMT
predicts $K/\mu = 0.71$, as mentioned earlier.  Moreover, the
effective Poisson ratio from simulations, $\nu \approx 0.27$ is in
excellent agreement with that of the experiment $\nu \approx
0.28$, but greatly differs from the theoretical prediction
$\nu_e=0.02$, Eq. (\ref{nu_e}).

\subsection{Role of transverse forces and rotations}

To understand why $\mu$ is overestimated by EMT we must examine
the role of transverse forces and rotations in the relaxation
process of the grains. These effects do not play any role in the
calculation of the bulk modulus. According to the EMT, the
transverse force $F_t$
 contributes only to the shear modulus and not to
the bulk modulus [see Eqs. (\ref{emt_k}), (\ref{emt_mu}) and
(\ref{emt_mu2})]. We are therefore motivated to examine the
behavior of the moduli as a function of the strength of the
transverse force. We replace the tangential stiffness $k_t$ in Eq.
(\ref{ft}) by $\alpha k_t$ and redefine the transverse force as

\begin{equation}
\Delta F_t=\alpha ~ k_t ~ (R \xi)^{1/2}~ \Delta s, \label{newft}
\end{equation}
$\alpha = 0$ is appropriate for frictionless coupling (perfect
slip), whereas $\alpha = 1$ describes the fully frictional result
(perfect stick) and corresponds to the results described so far.
To quantify the role of the transverse force on the elastic
moduli, we calculate $K(\alpha)$ and $\mu(\alpha)$ varying
$\alpha$ from 0 to 1, at a given pressure, $p=100$ KPa, low enough
so that the changing number of contacts does not play a role.

\begin{figure}
\centering {\resizebox{8.5cm}{!}{\includegraphics{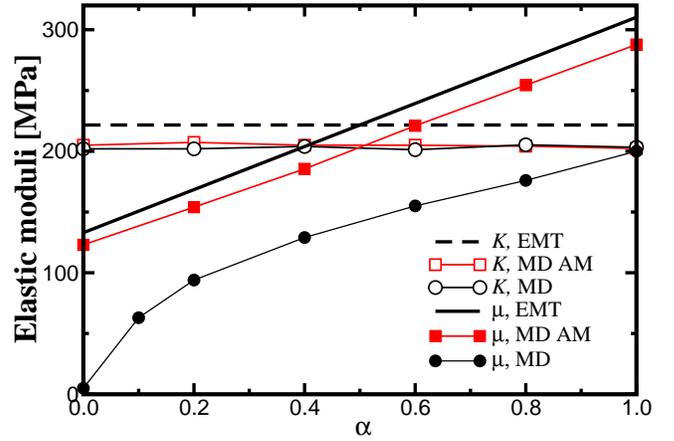}}}
\caption{$K(\alpha)$ and $\mu(\alpha)$ versus $\alpha$ for a fixed
$p=100$  KPa as calculated numerically with MD (noted MD), as
calculated numerically using only the affine motion (noted MD AM)
and as predicted by the EMT (noted EMT).} \label{alpha}
\end{figure}

The results are plotted in Fig. \ref{alpha} (curves labelled
MD). To compare with the theory we plot the prediction of the
EMT Eqs. (\ref{emt_k}) and (\ref{emt_mu2}) in which $k_t$ is
rescaled by $\alpha k_t$ (curves labelled EMT).  (The curves
labelled MD AM are discussed in the next subsection.)  The
simulation confirms that $K$ is essentially independent of the
strength of the tangential force; both theory and simulations show
a flat line in Fig. \ref{alpha}. Surprisingly, the shear modulus
is extremely sensitive to the tangential force and becomes
negligible small in the limit of frictionless particles ($\alpha
\rightarrow 0$) dropping to less than 10\% of the predicted EMT
value.  We see that the EMT badly fails in accounting for the
vanishing of the shear modulus as $\alpha\to0$. By contrast the
bulk modulus agrees reasonably well with EMT regardless of whether
there was perfect slip or perfect stick. What is the most serious
problem with the elastic theory? In the next section we will focus
on the role of stress relaxation and the nonaffine motion of
grains due to disorder.

First, however, we wish to eliminate a conceptually simpler effect
of disorder as the explanation for the behavior of $\mu(p)$.  In
the simulations (and presumably in the experiments), it is not
true that each grain has the same number of contacts.  Rather,
there is a distribution of contacts ranging from $Z=3$ to $Z= 10$
with a peak at $Z = 6$, which is near the average ($\bar{Z} =
6.14$ at 100 KPa). Thus, the local elasticity moduli can vary
widely from one grain to another.  There is a well-developed
theory for just such situations \cite {bema}, which is also called
a ``self-consistent effective medium approximation" (sc-ema). Let
$K_i$ and $\mu_i$ be the moduli for spherical inclusions whose
volume fraction is $c_i$. The effective elastic constants for the
composite, $K^*$ and $\mu^*$, are determined by the simultaneous
solution of the following coupled equations:
\begin{equation}
\sum_i c_i \frac{K^*-K_i}{K_i + (4/3)\mu^*} = 0, \label{Kema}
\end{equation}
and
\begin{equation}
\sum_i c_i \frac{\mu^*-\mu_i}{\mu_i + F^*} = 0, \label{mema}
\end{equation}
where \begin{equation} F^* =
\frac{\mu^*(9K^*+8\mu^*)}{6(K^*+2\mu^*)}\:\:\:.\label{Fema}
\end{equation}
Effective medium theories of this sort generally work well in
situations in which the disorder is not too great (such as when
there is a log-normal distribution of constituent properties, or
when one is near a percolation threshold).  Moreover, the sc-ema
have certain desired properties, such as correct limiting values
and lying within upper and lower bounds.  See \cite {bema} for
details.

Here we take the view that the system is a composite consisting of
spherical inclusions, each of which has moduli given by Eqs.
\rf{emt_k} and \rf{emt_mu2}.  In the case at hand it is useful to
rewrite them in terms of the local value of the compressive
strain, $\epsilon_i<0$, within each inclusion [See Ref.
\cite{johnson1} for details]:
\begin{equation}
K_i = \frac{\phi k_n}{12\pi}Z_i (-\epsilon_i)^{\frac{3}{2}},
\end{equation}
\begin{equation} \mu_i = \frac{\phi (k_n + \frac{3}{2} \alpha k_t)}{20\pi}Z_i
(-\epsilon_i)^{\frac{3}{2}}. \label{km_i}
\end{equation}

Of course, grains with a large number of contacts, $Z_i$, can be
expected to have a smaller than average compressive strain,
$\epsilon_i$.  In order to relate $\epsilon_i$ to the macroscopic
strain, $\epsilon^*$, we recognize that the spirit of the sc-ema
is that each spherical inclusion is surrounded by the host
material. Therefore, it is a simple elasticity problem to show
that the differential change in strain within a spherical
inclusion is related to the differential change in macroscopic
strain by
\begin{equation}
d \epsilon_i = \frac{K^* + (4/3)\mu^*}{K_i + (4/3)\mu^*} d
\epsilon^*\:\:\:.\label{inclusion}
\end{equation}
We take the distribution of contacts $\{c_i\}$ from our simulation
at 100 KPa.  It is straightforward to solve the system of equations
\rf{Kema}-\rf{inclusion}.  The EMT we have been discussing
corresponds to $Z_i \rightarrow <Z>$ and $\epsilon_i \rightarrow
\epsilon^*$; for the case $\alpha = 0$, and using the same
material parameters as before, it may be written as
\begin{equation}
K_e = 16.2 (-\epsilon^*)^{3/2},
\end{equation}
\begin{equation}
\mu_e = 9.7 (-\epsilon^*)^{3/2}, \label{old}
\end{equation}
where the moduli are expressed in GPa.  If, though, the full
distribution of contact numbers is used in the foregoing analysis
the results are
\begin{equation}
K^* = 15.8 (-\epsilon^*)^{3/2},
\end{equation}
\begin{equation}
\mu^* = 9.5 (-\epsilon^*)^{3/2} \:\:\:.\label{new}
\end{equation}

The point of this exercise is to demonstrate that, although the
packing is obviously disordered, the effect of the disorder alone
is quite negligible as far as the macroscopic elastic moduli are
concerned. Similar results hold for $\alpha = 1$.  Each grain
sees, more-or-less, the same average environment as any other.  In
the next section we investigate the effects of disorder induced
relaxation, which, we believe is the underlying effect behind the
small values of $\mu(p)$ we are observing.


\subsection{Role of relaxation and disorder}

In the EMT, we saw that if an affine  perturbation of the form
(\ref{affine}) is applied to the system, the grains are always at
equilibrium due to the assumption of isotropic distribution of
contacts and further relaxation of the grain is not significant.
The response is then purely elastic.

\begin{figure}
\centering {\resizebox{8.5cm}{!}{\includegraphics{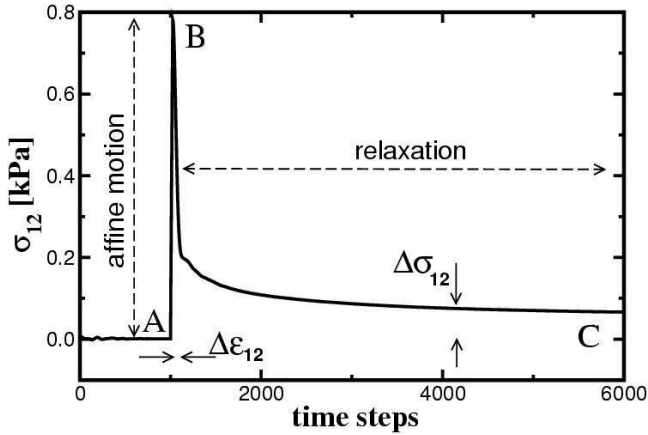}}}
\caption{Relaxation of the shear stress (B$\to$C) after an affine
motion (A$\to$B) in the calculation of the shear modulus.}
\label{nonaffine}
\end{figure}

On the contrary, in the MD simulations (and in the experiments)
after the application of an affine perturbation via the motion of
the boundaries and grains, the beads in the immediate neighborhood
of each grain move around, relative to the center grain, in a way
which gives rise to a stress relaxation associated with these
rearrangements of particles.

Figure \ref{nonaffine} shows the  behavior of
$\sigma_{12}(t)\equiv \epsilon_{12}G(t)$ as per Eq. \rf{fred} for
a system at $p=100$ KPa and with $\alpha=0.2$ during and after the
application of the affine strain perturbation
$\Delta\epsilon_{12}$ which moves all the grains according to the
external strain Eq. (\ref{affine}). We see how the system behaves
as a viscoelastic solid as explained in Section
\ref{viscoelastic}. When the affine perturbation is applied, the
shear stress increases (from A to B in Fig. \ref{nonaffine}) and
the grains are far from equilibrium since the system is
disordered. This is the instantaneous elastic response.
 The grains then relax towards equilibrium as 
(from
B to C), and we measure the resulting change in stress
$\Delta\sigma_{12}$ as $t \to \infty$ 
from which the modulus $\mu$ is calculated
as in Eq. (\ref{mu1}).

For a better understanding of the approximations involved in the
EMT, suppose we repeat the MD calculations now taking into account
only the affine motion of the grains and ignoring the subsequent
relaxation. The resulting values of the moduli are obtained as
$\mu_{\mbox{\scriptsize affine}} = \Delta\sigma_{12}
^{\mbox{\scriptsize affine}} /\Delta\epsilon_{12} $ with
$\Delta\sigma_{12} ^{\mbox{\scriptsize affine}}$ defined in Fig.
\ref{nonaffine}. In Fig. \ref{alpha} we plot the moduli calculated
in this way as a function of $\alpha$ for $p=100$ KPa (curves
labelled MD AM). The affine moduli are very close to the EMT
predictions: there remains a 10\% difference between the EMT and
the MD (affine) which is representative of the disordered packing
which is averaged in the EMT. Thus, the difference between the MD
and EMT results for the shear modulus lies mostly in the
non-affine relaxation of the grains; this difference is largest
when there is no transverse force.

By contrast, grain relaxation after an applied compressional
affine perturbation is not particularly significant and the EMT
predictions for the bulk modulus are quite accurate as seen in Fig. 
\ref{alpha}.

\subsection{The isostatic limit as a critical point}

The surprisingly small values we found for $\mu$ as $\alpha
\rightarrow 0$ raises several questions.  We notice that $k_n$ and
$p$ are the only variables with the dimension of pressure in this
limit.  A scaling argument would lead to
\begin{equation} \mu \sim k_n (p/k_n)^\eta.
\label{new_scaling}
\end{equation}
The Hertz theory predicts $\eta=1/3$, a result which we find to be
valid at low pressure for frictional grains.  Indeed, quite
generally if one assumes that each grain-grain force scales as
Eqs. \rf{fn} and \rf{ft} and if one assumes the arrangement of the
grains, however disordered that may be, does not change with
pressure then both moduli scale as in Eq. \rf{new_scaling} with
$\eta = 1/3$.  This argument specifically presupposes that e.g.
the average coordination number does not change with pressure.

\begin{figure}
\centering {\resizebox{8.5cm}{!}{\includegraphics{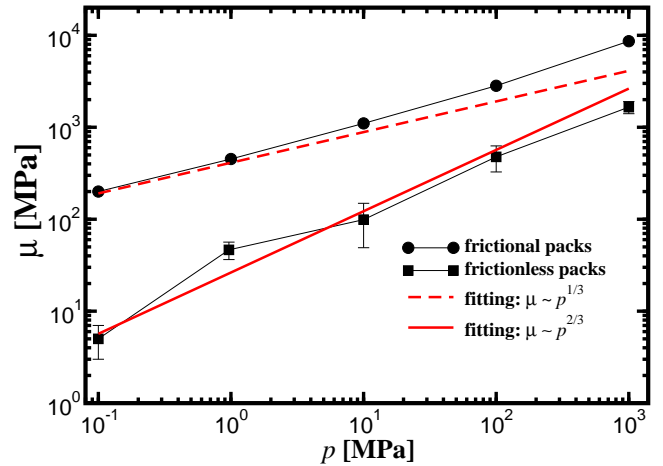}}}
\caption{Shear modulus versus pressure for frictional ($\alpha=1$)
and frictionless ($\alpha=0$) particles.}
\label{gamma}
\end{figure}

Since there are no other constants that could reduce the value of
$\mu$ for $\alpha \rightarrow 0$ we are lead to believe that a new
exponent $\eta$ should describe the shear modulus for frictionless
packs. This is an effect which lies outside the standard
assumptions of elasticity theory, as indicated above.  Since
$p<k_n$, then $\eta>1/3$. To give validity to our hypothesis, we
plot in Fig. \ref{gamma} $\mu(p)$ for $\alpha=1$ and $\alpha=0$.
We see that a better fit to the low pressure behavior of $\mu(p)$
for $\alpha=0$ is achieved with $\eta=2/3$. Notice that we
deliberately try to fit the data at low pressure to avoid the
issue of the increasing coordination number.

How can we explain the $2/3$ scaling behavior? A possible answer
could be provided by a recent conjecture by Alexander
\cite{alexander} who proposed the following scaling
\begin{equation}
\mu \sim k_n {\cal A} (p) (p/k_n)^\eta. \label{new_scaling2}
\end{equation}
where the function ${\cal A}(p)$ is determined by the geometry of
the reference frame of rigidity, Eq. (\ref{R}), which is
determined, in turn, by the pressure. Assuming that the limit $p
\to 0$ is indeed a critical state of rigidity, then we expect
\begin{equation}
{\cal  A}(p) \sim p^\lambda , 
\end{equation}
which would explain the anomalous scaling for the frictionless
grains with  $\lambda = 1/3$, while for frictional grains we would
have $\lambda=0$.



\subsection{Microstructure and  force chains.}

The velocity of acoustic signals probes an effective medium which
should be
 homogeneous at length scales larger than a typical
correlation length of the material.
Experimental and numerical work
indicates that there is an internal structure at length scales
$\sim10d$, where $d$ is the typical size of the grains: the
forces are observed to be localized along ``force chains''
carrying most of the loads in the system (see Fig. \ref{chains})
\cite{makse-fc,fc3,cundall2,chicago1}.  A question of interest is
how such a microstructure affects the properties of the system at
macroscopic length scales where the elastic continuum theory is
valid \cite{jussieu}.

We want to quantify the relevance of force chains to the elastic
moduli. We calculate the shear modulus as a function of a subset
of forces belonging to the strongest forces in order to search for
the backbone of grains which give rise to the shear rigidity of
the material. Is this backbone determined by the force chains, or
do the interstitial particles play also a relevant role to
determine the rigidity?

In this regard, recent calculations of Radjai {\it et al.}
\cite{radjai2} have shown that the stress ratio between shear and
compression shows a ``percolation-like'' behavior: the forces
larger than the average are responsible for most of the rigidity
of the material. This was shown to be valid in 2D. Here we follow
\cite{radjai2} and define a $\zeta$-network which includes only
forces smaller than a cutoff force $\zeta$. Then we redefine the
stress Eq. (\ref{stress}) and compute the shear stress only for
the $\zeta$-network as
\begin{equation}
\sigma_{12}(\zeta)=\frac{1}{2 V}\sum_{|\mbox{\bf F}| <\zeta}(F_{i}
R n_j + R n_i F_j), \label{stress2}
\end{equation}
from which we obtain the shear modulus for the $\zeta$-network as
$\mu(\zeta) = \Delta\sigma_{12}(\zeta)/\Delta \epsilon_{12}$ (for
$\zeta\to \infty$ we recover our previous results).

\begin{figure}
\centering {\resizebox{8.5cm}{!}{\includegraphics{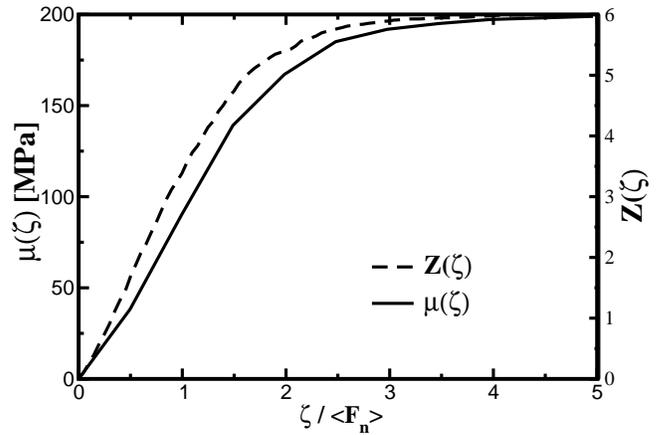}}}
\caption{ Behavior of the shear modulus and the coordination number for
the $\zeta$-network. We use the packing at 100 KPa depicted in
Fig. \protect\ref{chains}a for this calculation.} \label{zeta}
\end{figure}

\begin{figure}
\centering {(a) \resizebox{8cm}{!}{\includegraphics{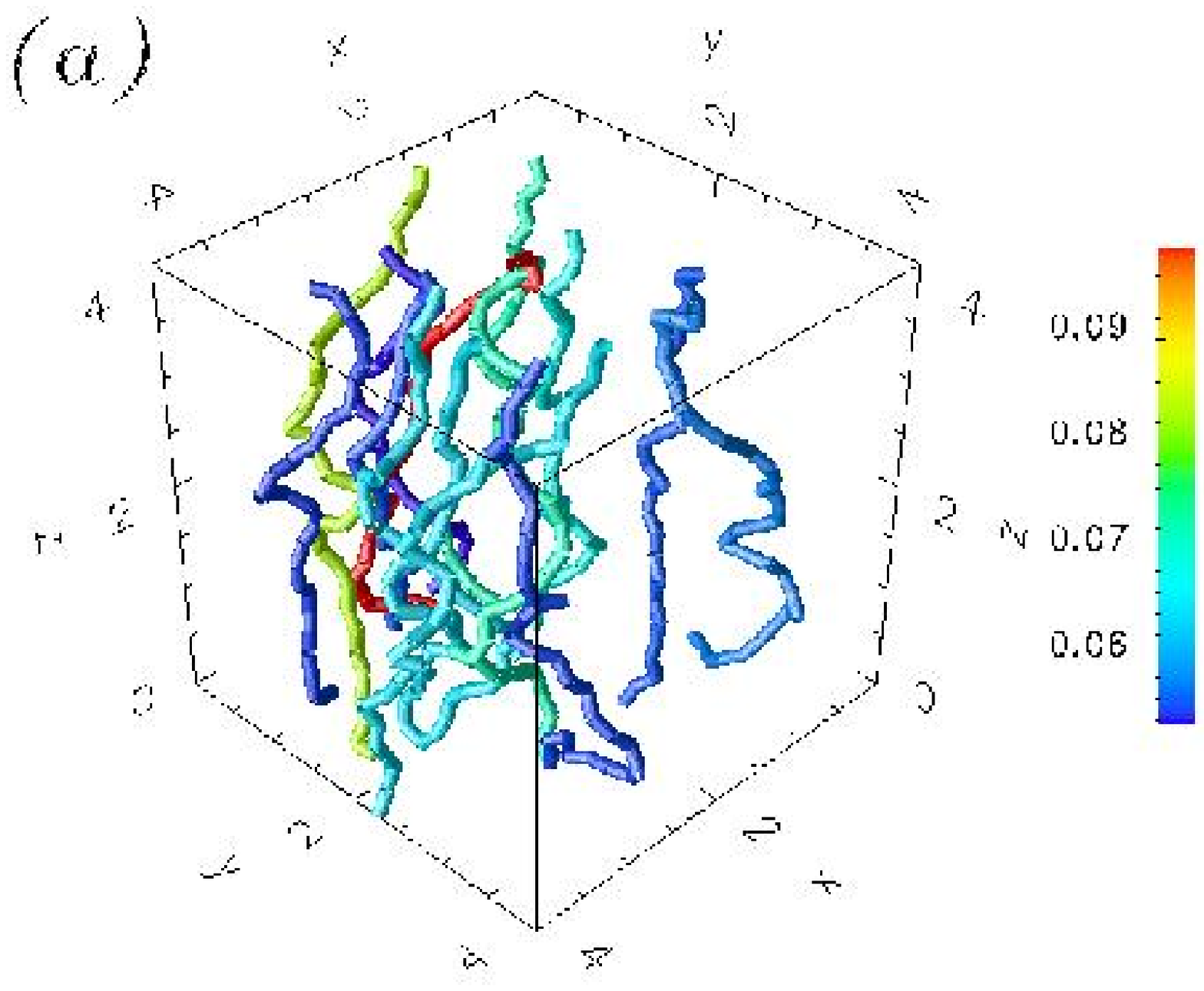}}}
\centering {(b) \resizebox{8cm}{!}{\includegraphics{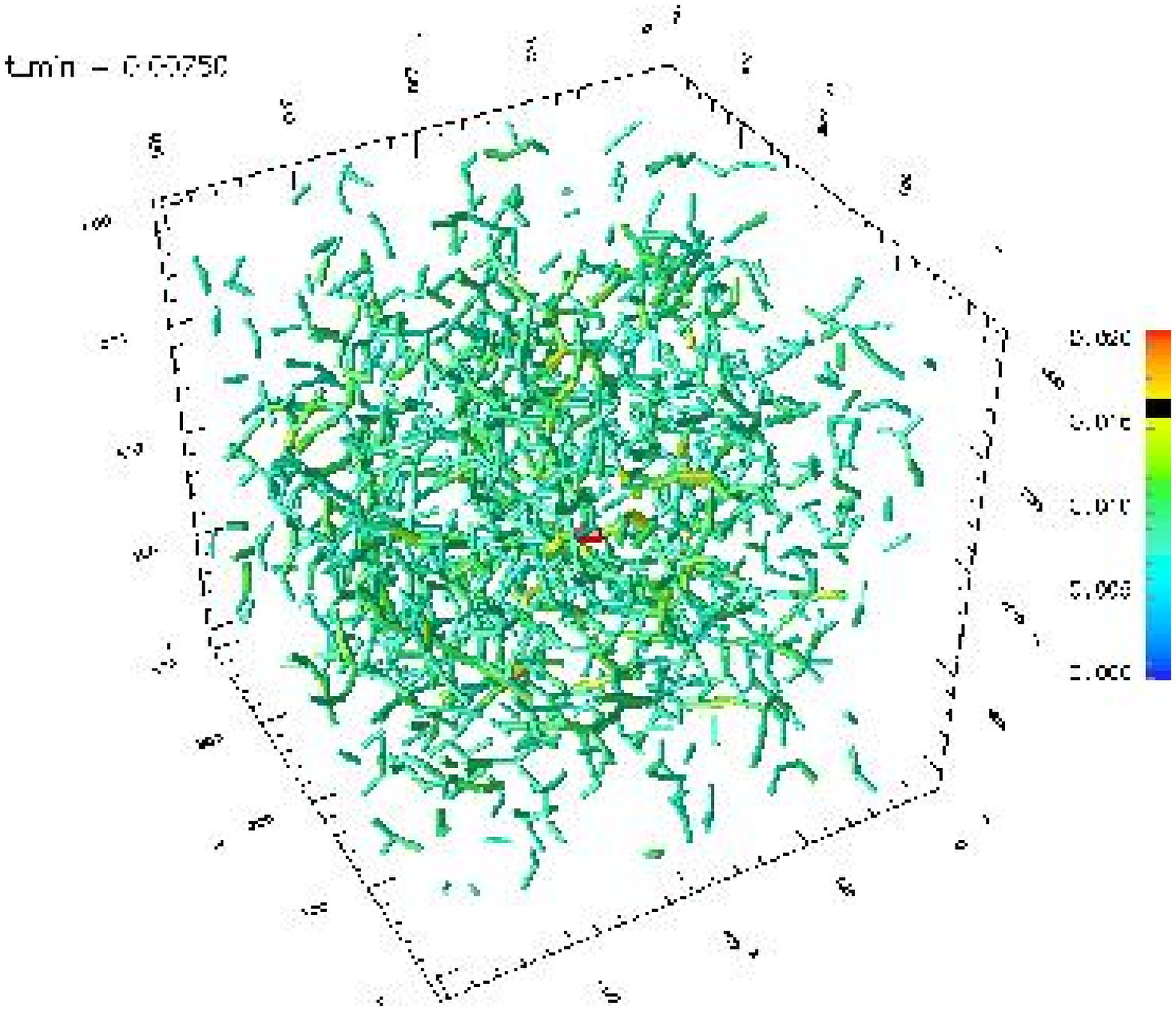}}}
\centering {(c) \resizebox{7cm}{!}{\includegraphics{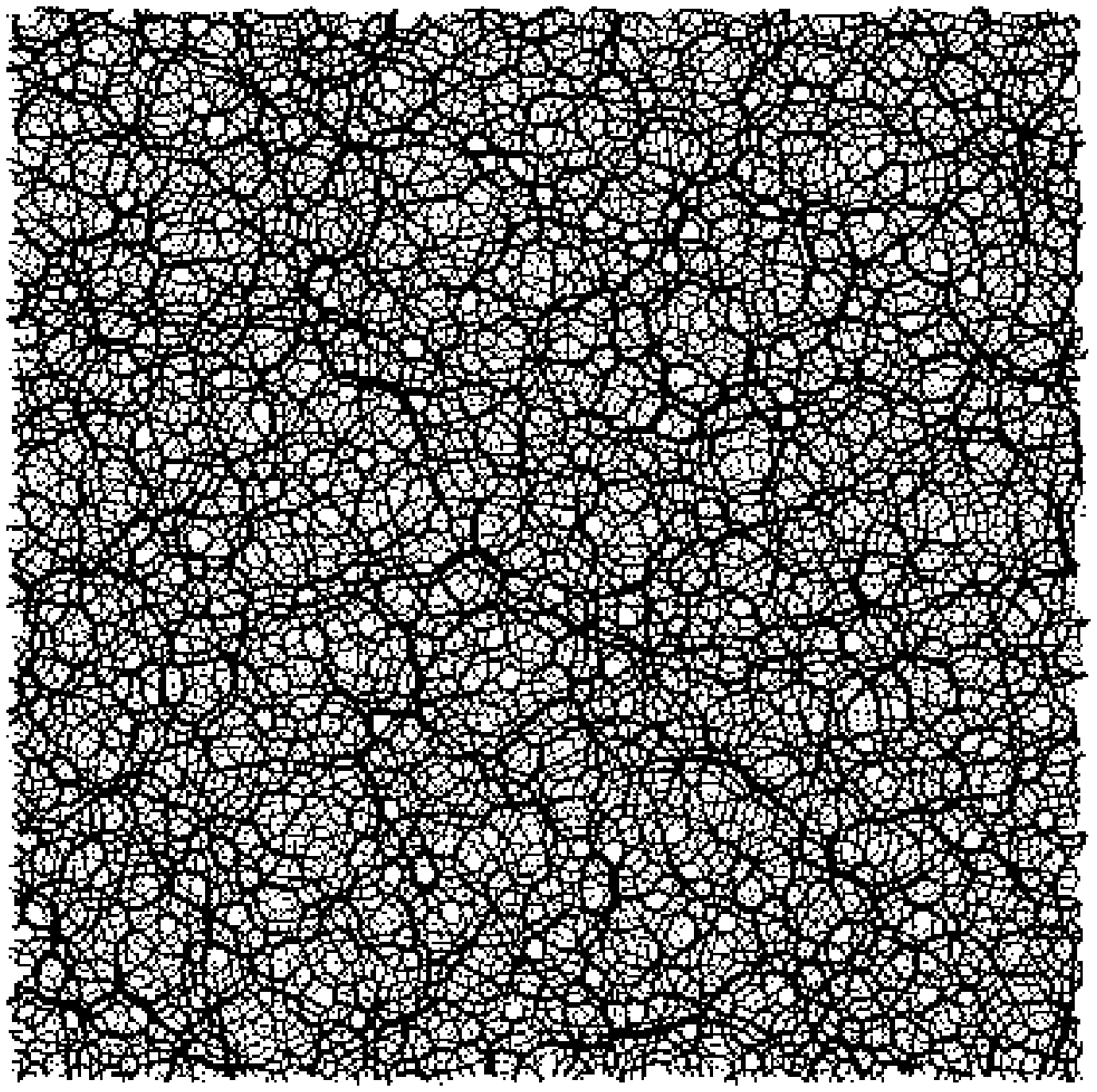}}}
\caption{(Color online) Force chains in granular matter: (a)
Frictional system under
uniaxial compression near $\phi_c$ (from \protect\cite{makse-fc}).
Percolating force chains are seen in this case. We apply an
algorithm which  looks for force chains by starting from a sphere
at the top of the system, and following the path of maximum
contact force at every grain.  We plot only  the paths which
percolate, i.e., stress paths spanning the sample from the top to
the bottom. (b) Frictionless
isotropic system at $p=100$ KPa in 3D. We plot only the forces
larger than the average. Force chains seem
to be tenuous and not well defined. 
(c) Force chains in a 2D frictional system. Force
chains are clear in this case.} \label{chains}
\end{figure}

Figure \ref{zeta} shows the result of $\mu(\zeta)$ for $p=100$ KPa
and $\alpha=1$ and should be compared with 
Fig. 4 in \cite{radjai2}.  
In contrast with the 2D results of \cite{radjai2}
we find no evidence of a bimodal distribution of forces which
would give rise to a percolation-like behavior of the shear
modulus. We see that the shear modulus and the coordination number
increase continuously as we increase $\zeta$.

We also repeat the same calculations for our two-dimensional
packings and find the same result as in 3D, i.e. we find no evidence of
a  bimodal character in the behavior of
the shear modulus versus the force cut-off.
The fact that we do not see the same behavior in 2D as in \cite{radjai2}
might be related to the regularization scheme used in our MD simulations
to handle the frictional forces which may eliminate the critical
behavior found in \cite{radjai2}.  Radjai {\em et al.} 
used a Contact Dynamics algorithm 
which tackle the non-smooth character of the interactions without
any regularization schemes.

Figure \ref{chains} shows our attempt to visualize 
force chains in 3D packings (a) without friction under isotropic compression, 
(b) with friction under uniaxial
compression  
and (c) in 2D frictional isotropic packings.
Force chains are not prominent in the 3D isotropic frictionless packing.
Moreover, the continuous variation of $\mu(\zeta)$ obtained for this packing
seems to
indicate that all forces are important for the mechanical response
to shear, and not just the larger forces which may be organized in
force chains.
However, force chains are prominent in 
the 3D packing under uniaxial compression and the 2D packing.

\section{Theory: single particle relaxation}
\label{sectheory}

Since the difficulty with the shear modulus is shown to be due to
the relaxation of the particles from the initial uniform strain
approximation, we next perform the simplest investigation that
allows for some relaxation.  From the simulations, we know the
rest positions of each of the particles, as well as the contact
vectors $\hat{\bf d}^{(q)} = ({\bf x}_1-{\bf x}_2) / | {\bf
x}_1-{\bf x}_2|$ (the vector from a particle to each of the
particles with which it is in contact). Consider a specific
particle. We make the approximation that when a small amplitude
macroscopic strain is applied its contacting particles move
according to the affine approximation. The particle will
experience an unbalanced force and an unbalanced torque.
Accordingly, it will relax to a new position and orientation such
that the net force and torque on it become zero. So, for the
specific particle we calculate its new position and orientation.
We next calculate the energy stored within each of the contact
``springs". We do this for each of the particles in the simulation
to calculate the total stored energy due to the applied strain and
we set this equal to the usual expression for strain energy in
order to deduce the new estimates for the bulk and shear moduli of
the aggregate. This procedure is detailed below.

Consider a particle, labeled $a$, which we take to be centered at
the origin.  It has $z_a$ contacts at the positions $\{{\bf
d}^{(q)}: q=1,z_a\}$.  Assuming that one of the contact points is
displaced by an amount ${\bf u}^{(q)}$ the increment in the
intergrain force at contact $q$ is

\begin{equation}
{\bf F}^{(q)}_u = K_N [(\hat{\bf d}^{(q)}\hat{\bf d}^{(q)}) \cdot
{\bf u}^{(q)} ] + \alpha K_T [( {\bf I} - \hat{\bf
d}^{(q)}\hat{\bf d}^{(q)}) \cdot {\bf u}^{(q)}], \label{D1}
\end{equation}
where $K_N$ and $K_T$ are given by:
\begin{equation} K_N = \frac{ 2 \mu_g R^{1/2}}{1- \nu_g } \xi^{1/2},
\end{equation}
\begin{equation} K_T = \frac{ 4 \mu_g R^{1/2}}{2- \nu_g }
\xi^{1/2},
\end{equation}
and $\xi$ is the normal displacement which can be related to the
external pressure through the average affine approximation
\cite{walton} by
\begin{equation}
\xi = R  \left [ \frac{3 \pi}{2} \frac{(1-\nu_g)}{\phi Z}
\frac{p}{\mu_g} \right ]^{2/3}.
\end{equation}
The parameter $\alpha$ allows us to continuously investigate the
crossover behavior from perfect slip ($\alpha = 0$) to perfect
stick ($\alpha = 1$).

As written, the total force on the specific particle, due to the
sum of all the contact forces is not zero: \begin{equation} {\bf
F}_u \stackrel{def}{=} \sum_q {\bf F}^{(q)}_u \neq 0
\:\:\:.\end{equation} Accordingly, that particle will move to a
new equilibrium position, ${\bf X}$.  Similarly, the net torque on
the particle is unbalanced:

\begin{equation}  {\bf N}_u \stackrel{def}{=} \sum_q  {\bf
d}^{(q)} \times  {\bf F}^{(q)}_u \neq 0.
\end{equation}

Accordingly, the particle will rotate through an angle $\bfomega$
to a new orientation.  The generalization of Eq. (\ref{D1}) that
takes into account the new position and orientation is

\begin{equation}
\begin{array}{ll}
 {\bf
F}^{(q)}= & K_N [(\hat{\bf d}^{(q)}\hat{\bf d}^{(q)}) \cdot ({\bf
u}^{(q)}-{\bf X}) ] + \\
& \alpha K_T [( {\bf I} - \hat{\bf d}^{(q)}\hat{\bf d}^{(q)})
\cdot ({\bf u}^{(q)}-{\bf X})- \bfomega\times{\bf d}^{(q)}],
\end{array}
\label{D6}
\end{equation}

Now, the requirement that the particle is in equilibrium with its
contact forces, $\sum_q {\bf F}^{(q)} \stackrel{set}{=} 0$, gives
three linear equations in the six unknowns, $\bfomega$ and ${\bf
X}$. The requirement that the total torque must vanish, $\sum_q
{\bf d}^{(q)} \times  {\bf F}^{(q)} \stackrel {set}{=} 0$, gives
the remaining three.  It is straightforward to solve these
equations numerically.

Having determined the new equilibrium position and orientation,
one can show that the total work done by the contact forces on the
$a$-th particle is simply
\begin{equation}
\begin{array}{ll}
W_a = & \frac{1}{2} \{
K_N \sum_{q=1}^{z_a}(\hat{\bf d}^{(q)} \cdot {\bf u}^{(q)})^2  + \\
 &
\alpha K_T  \sum_{q=1}^{z_a}\mid \hat{\bf d}^{(q)} \times {\bf
u}^{(q)}\mid^2 - {\bf F}_u \cdot {\bf X} -  {\bf N}_u \cdot
\bfomega  \},
\end{array}
\label{D7}
\end{equation}
$\bf{X}$ and $\bfomega$ are determined as described above.  In
order to calculate $W_a$ we make the affine assumption, that the
displacement at the contact point is simply related to the
macroscopic strain by Eq. (\ref{affine}).  Since we know the exact
positions of each contact vector, ${\bf d}^q$, from the
simulations, we are able to evaluate Eq. \rf{D7} for each particle
in the ensemble.

We now evaluate $ \sum_a W_a/V$  for a pure compression and for a
simple shear numerically and we equate the result to the elastic
energy, Eq. (\ref{energy}), in order to deduce the values of $K$
and $\mu$.



The above procedure can only reduce the moduli relative to those
of the effective medium prediction. If, in Eq. \rf{D7}, we assume
there is no relaxation [$\bfomega = 0$ and $\bf{X} = 0$], and if
we replace the sum over contacts by an integral over a presumed
uniform distribution of contact directions, we reproduce the
effective medium theory, Eqs. \rf{emt_k} and \rf{emt_mu2}.

\begin{figure}
\centering {\resizebox{8.5cm}{!}{\includegraphics{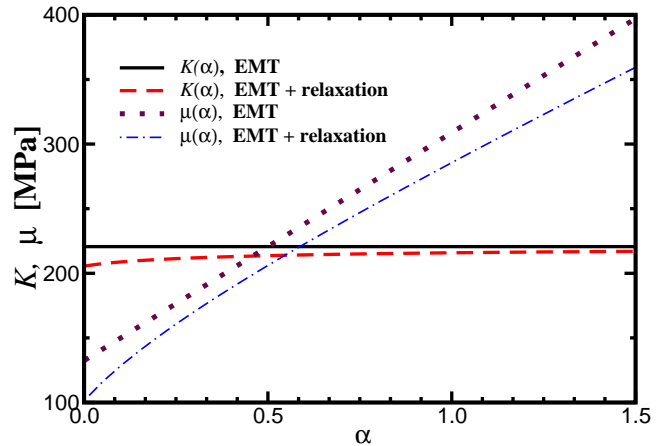}}}
\caption{First order correction to EMT allowing relaxation of
grains from the affine motion. This figure should be compared
with Fig. \protect\ref{alpha}.} \label{theory}
\end{figure}

The results of such a calculation are shown in Fig. \ref{theory},
which is to be compared to Fig. \ref{alpha}.  The static confining
pressure is 100 KPa.  We see that, relative to the effective
medium prediction, there is a small reduction of the bulk modulus,
which is relatively insensitive to $\alpha$. There is a much
larger reduction of the shear modulus but the results of the
simulations for the shear modulus give values that are even
smaller still.  For $\alpha=0$ (perfect slip) the simulations give
$\mu = 8 \pm \: 3$ MPa, which is essentially indistinguishable
from zero, whereas from Figure \ref{theory} we have a value of 100
MPa.  We see that relaxation effects at the single particle level,
while significant, are by no means sufficient to explain the
effect.  In the fully frictional case of $\alpha = 1$ there is a
reduction relative to the EMT but the simulation gives a value of
$200 \pm \: 10$ MPa. (In Fig. \rf{theory} we have extended the
calculations into the unphysical range of $\alpha > 1$ to
emphasize that there is a slight change of slope, relative to the
EMT.)

We are thus lead to consider a more sophisticated theory in which
we explicitly account for collective fluctuations. The next step
in this direction is developed in  \cite{laragione} where we
introduce fluctuations in pairs of contacting particles.  This
theory is developed for  the frictionless case, where the
reduction in shear modulus is most dramatic and for which we can
derive an analytic result using some fairly weak assumptions.

\section{Summary and outlook}
\label{summary}

Where do we go from here? We clearly need new theoretical
frameworks to describe the collective relaxation  of granular
materials, especially under shear and for frictionless packs.
Below we give a short review of some of the ideas that have been
proposed recently, and how these theories are related to our
results.

\subsection{Elastic versus fragile matter}

We have seen that the impossibility of defining a strain field
which is inhomogeneous at the level of the grain is at the root of
the problems of the elastic theory: the EMT approach relies on the
assumption of a uniform strain field  at all scales \cite{landau,green3}.

Interestingly, recent studies
\cite{edwards-grinev,fragile,ball} have proposed theories
of stress transmission in granular packs which describe the
internal stresses without resorting to the use of strain
variables, as in elasticity theory. These groups argue that
cohesionless grains are in a ``fragile state'' of marginal
rigidity or isostatic at a minimal coordination number $Z_c$ and
they are only able to support certain loads without severe
rearrangements. A novel closure relation between stress
components--- for instance, the fixed principal axis ansatz--- and
not between stress and strain--- as in elasticity--- has been
proposed to solve for the indeterminacy in the granular system
\cite{fragile}.


The correct type of closure relation (elastic or fragile) is still
a question of much debate \cite{savage-elastic}, although there are
recent experiments on the single-particle Green function
measurements suggesting that the elastic framework might be the
correct approach at large scales \cite{green1,green2,green3}.

In the case of collective relaxation dynamics, our results show
that the elastic formulation is erroneous in describing the
macroscopic shear response of granular materials. Moreover, we
find that a very small shear modulus appears for frictionless
packs. This shear modulus decreases towards zero as $p\to 0$, as
$\phi\to \phi_{\mbox{\scriptsize RCP}}$, and as the system
approaches the isostatic limit of $Z\to Z_c=6$.

The vanishing of the shear modulus could be interpreted as a
``fragile'' behavior. In the limit $\alpha \to 0$ a packing of
nearly rigid particles responds to an external isotropic load with
an elastic deformation and a finite $K$, since the external
perturbation is compatible with the principal axes of the stress
predetermined by the preparation history of the sample. By
contrast, such a system cannot support a shear load ($\mu \to 0$)
without severe particle rearrangements. Thus the granular system
supports, elastically, only perturbations compatible with the
structure of  force chains and deform irreversibly otherwise, i.e.
it is in a ``fragile'' state.

\subsection{Jamming and melting}

Our results show that the fragile limit is approached as the
system gets closer to RCP limit, and  that at RCP there is a
jamming transition between a liquid-like state and a solid-like
state with a finite modulus. The approach to the critical point is
characterized by several power-law exponents as in a second-order
phase transition. The vanishing of the shear modulus can be
understood as a melting of the system occurring when the system
approaches the isostatic point. This fluid like behavior has
similarities with melting transitions found in compressed
emulsions, and foams \cite{lacasse,durian,bolton} near the RCP
fraction. A slow relaxation time and the increase of the
correlation length between force chains is found near RCP. This
behavior indicates that the physics of granular materials might be
closely related to other complex  systems undergoing jamming as
proposed recently \cite{liu} such as glasses, colloids, foams, and
emulsions.

\subsection{Conclusions}

Our MD simulations are in good agreement with the available
experimental data on the pressure dependence of the elastic moduli
of granular packings. They also serve to clarify the deficiencies
of EMT.  Grain relaxation after an infinitesimal affine strain
transformation is an essential component of the shear (but not the
bulk) modulus. This relaxation is not taken into account in the
EMT.

Clearly, there is a need for alternative theories to describe
granular packings. Recent work on stress transmission in minimally
connected networks may provide an alternative formulation and
allow a proper description of the response of granular materials
to external perturbations.

Acknowledgments: This work was supported by the DOE, Chemical Sciences,
Geosciences and Biosciences Division, and the NSF, Division of Materials
Research.


\section{Appendix}

{\bf Appendix A: Resistance against rolling and tangential force
with microslip}
\renewcommand{\theequation}{A.\arabic{equation}}
\setcounter{equation}{0}

Our model of the intergrain contact is based on two assumptions.
First, we consider the no-slip solution of Mindlin for the
tangential force, and we consider total slip of the contact area
only when the total tangential force exceed $\mu_f F_n$. However,
in reality, the contact may slip over an annular ring of the
contact area for any finite value of the tangential force. A
general study for several loading histories considering that
microslip occurs, i.e., $|\Delta F_t| > \mu_f \Delta F_n$, was
performed by Mindlin-Deresiewicz \cite{dere}
 and analyzed in more detail by Thornton and Randall \cite{randall}.
They showed that the incremental tangential force can be obtained
as: $\Delta F_t= \varepsilon  k_t (R \xi)^{1/2} \Delta s
\pm \mu_f (1-\varepsilon) \Delta F_n,$ where $\varepsilon=1$ when
microslip does not occur ($|\Delta F_t| < \mu_f \Delta F_n$) and
$\varepsilon$ takes different values depending on the path loading
history of loading, unloading and reloading \cite{randall}. We
have done preliminary tests using this more general solution of
the tangential force, and found no significant changes in
comparison with the results obtained with the no-slip solution of
Mindlin. Therefore, we have performed our simulations using the
simpler Mindlin contact theory. Besides, the EMT calculations are
done using Hertz-Mindlin forces, so that we want to use the same
interparticle laws for a better comparison between numerics and
theory.

Second, while rotation of spherical grains is allowed in the
simulations, it is customary to model rotations without
resistance against rolling at the contacts \cite{cundall}.
Regarding this approximation, it should be pointed out that some
recent studies \cite{oda} showed that resistance against rolling
(modeled as an elastic spring yielding rotational resistance $k_r
\theta_r$, where $k_r$ is the rotational stiffness, and $\theta_r$
is the relative rotation by rolling) might be relevant for
modeling shear bands. The relevancy of rotational resistance to
static packings  has not been determined yet, and therefore, we do
not include it in our studies. It should be noted, however, that
the simulations consider resistance against shear given by the
elastic tangential force of Mindlin.

\vspace{.5cm}

{\bf Appendix B: Damping}

\renewcommand{\theequation}{B.\arabic{equation}}

\vspace{.5cm}

 Recently it has been shown that in order
to incorporate the dissipation law leading to inelasticity at the
grain-grain contact consistent with the Hertz contact law, a
nonlinear force dependency on the relative velocity of the grains
in contact has to be incorporated into the contact law
\cite{poschel2}.

This dissipative part of the normal force has been determined
recently by Brilliantov {\it et al.} \cite{poschel2} as:
\begin{equation}
F_n^{diss} = \frac{2}{3}~ A k_n R^{1/2}\xi^{1/2}
\dot{\xi}\:\:\:,\label{dissip}
\end{equation}
where $A$ is a relaxation time that depends on the viscous
properties of the grain material, and it can be uniquely
determined from experimental measurements of the coefficient of
restitution for spherical beads \cite{wolf-md,resti,bridges}.

In our studies, we are not interested in the way the system
approaches the equilibrium state, but only in the final state
which is supposed to be independent on the type of damping used.
Thus, we use the more efficient global damping and linear contact
damping described in Section \ref{md}. However, for dynamical
studies a damping term as in   Eq. (\ref{dissip}) should be
considered as well.

\vspace{.5cm}

{\bf Appendix C: Model of interaction between droplets}

\renewcommand{\theequation}{C.\arabic{equation}}

\vspace{.5cm}

In the case of emulsions, interdroplet forces are not given in
terms of bulk elasticity as in Hertz theory. Instead, forces are
given by the principles of interfacial mechanics without
considering shear forces
\cite{lacasse,durian,faraday,Princen1983}. For small deformations
with respect to the droplet surface area, the energy of the
applied stress is presumed to be stored in the deformation of the
surface. Hence, at the microscopic level, two spherical droplets
in contact interact with a normal repulsive force $F_n \sim R
\gamma  A$. This is the so-called Princen model
\cite{Princen1983}, where $A$ is the area of deformation, and
$\gamma$ is the interfacial tension of the droplets, and $R$ is
the geometric mean of the radii of the undeformed droplets. Since
the area of deformation is proportional to overlap $\xi$, then the
interdroplet interaction is $F_n\sim \gamma \xi$.

There have been more detailed numerical simulations
\cite{lacasse} to improve on this model and allow for
anharmonicity in the droplet response by also taking into
consideration the number of contacts by which the droplet is
confined. Typically these improved models lead to a force law for
small deformations of the form $F_n \propto A^b$ , where $A$ is
the area of deformation and $b$ is a coordination number
dependent exponent ranging from 1 (Princen model) to 3/2 (Hertz
model) (see also \cite{unknownauthors}).
%

\vspace{.5cm}

{\bf Appendix D: Time step}

\renewcommand{\theequation}{D.\arabic{equation}}

\vspace{.5cm}

The time step is usually chosen  much smaller than the collision
time. However, since each contact is enduring, the collision time
is extremely large and other conditions must be used. Besides, the
collision time for Hertz spheres depends on the relative
velocities of the particle, thus it does not defined a fixed time
scale \cite{landau}.

We choose the time step to be a fraction of the time that it takes
for a sound wave to propagate on the grain. Moreover, the
quasi-static approximation used to calculate the Hertz force is
valid only when the relative velocities of the particles is
smaller than the speed of sound in the grains \cite{poschel2}.
Thus, the  characteristic time is $t_0= R \sqrt{\rho_g/\mu_g}$.
Typically, one chooses a time interval much smaller than the
characteristic time,  then $\Delta t = a R \sqrt{\rho_g/\mu_g}$
with $a<1$. Typical values for glass beads are: $\rho =
2600$ Kg/m$^3$,  $\mu_g \approx 29$ GPa, $R=0.1$ mm. Then 
$\Delta t$ should be smaller than $10^{-8}$ s. Thus, in order to
perform a simulation over one  second, more than $10^8$ 
MD steps are needed, which is obviously a 
very intensive computation. In this case, it is customary
to increase the density or decrease the  
rigidity of the particles to allow for a
larger time step to integrate the equations of motion
over realistic periods of time. 
If the shear modulus of the grains in decreased, then it should be checked
that the resulting stresses are several order
of magnitude smaller than $\mu_g$, thus ensuring the condition 
of a nearly rigid system even though $\mu_g$ is taken smaller
to obtain larger time steps.

\vspace{.5cm}

{\bf Appendix E: Results in 2D}

\vspace{.5cm}

Here we show the results for the bulk and shear modulus
as a function of the pressure for a two-dimensional pack of 
spherical particles interacting via Hertz-Mindlin forces.
The 2D simulations are done with spherical Hertz-Mindlin balls
constraint to move in a plane. Thus the interparticle force
is that of the 3D case. Our system is not the same as a 
packing of disks in 2D since  the latter has a different
interaction law between particles.
Our results are  analogous to the three-dimensional case
shown in Fig. \ref{mu_1/3}.
All the conclusions regarding the moduli 
obtained for 3D are valid in this case as well.

\begin{figure}
\centering {\resizebox{8.5cm}{!}{\includegraphics{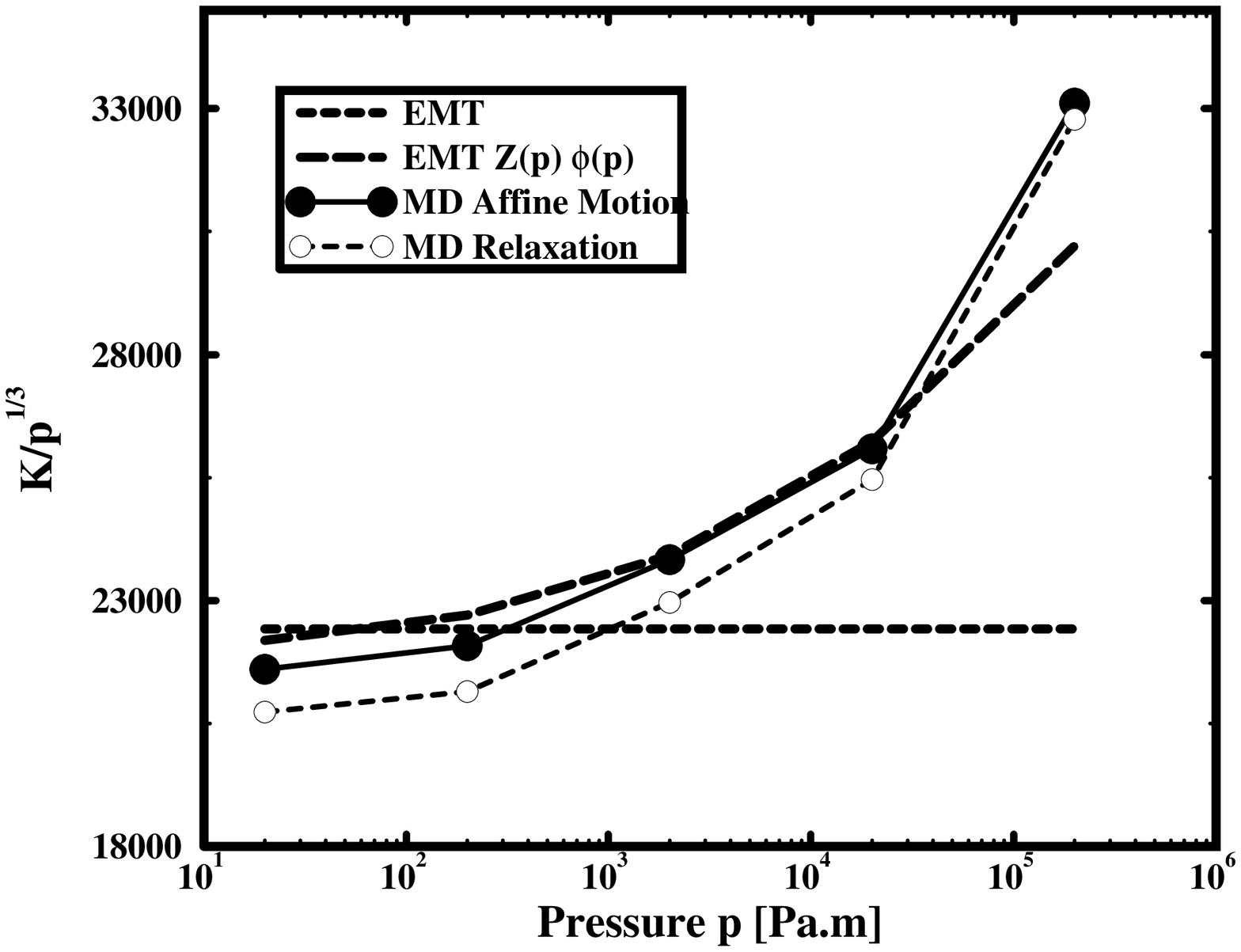}}}
\centering {\resizebox{8.5cm}{!}{\includegraphics{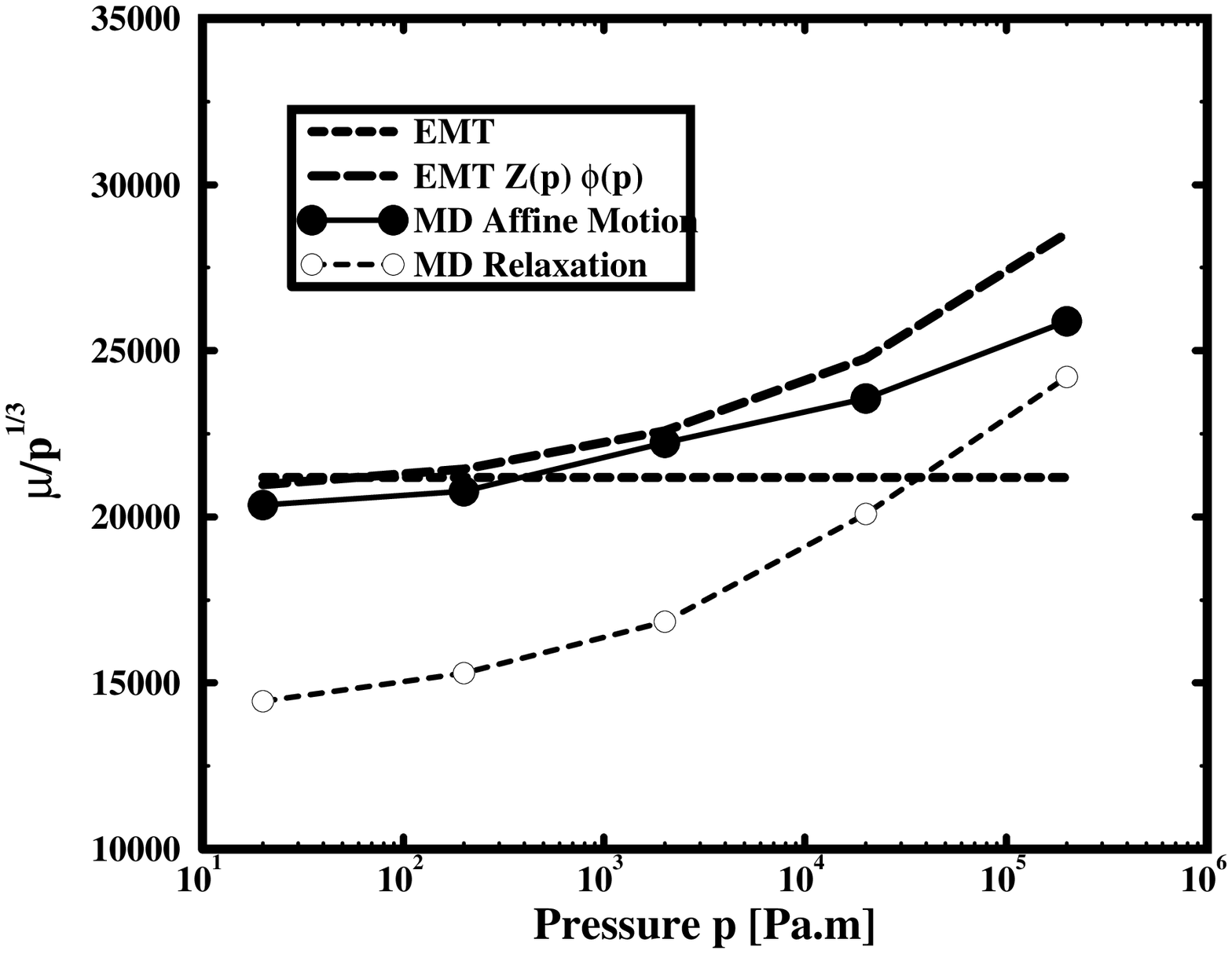}}}
\caption{Bulk and shear moduli for a 2D packing normalized to
$p^{1/3}$, EMT and  corrected EMT taking into account the pressure
dependence of $Z(p)$ from Fig. \ref{coord}c as well as
$\phi(p)$ (see Eqs. (\protect\ref{z2d}) and (\protect\ref{phi2d}).}
\end{figure}

The scaling of the coordination number is  similar to the 3D case:

\begin{equation}
Z(p) = Z_c + \left(\frac{p}{ ~\mbox{18 KPa m}}\right)^{0.28(7)},
\label{z2d}
\end{equation}
with 
$Z_c \approx 4$.
 
For the volume fraction we obtain

\begin{equation}
\phi(p) = \phi_c + \left(\frac{p}{ ~\mbox{32
 MPa m}}\right)^{0.4(1)}.
\label{phi2d}
\end{equation}
with a critical value of $\phi_c\approx 0.835$,
which is the RCP limit in 2D. This latter exponent is 
in disagreement with a mean field prediction based on the
contact law, which would imply an exponent 2/3 [see discussion after
Eq. (\ref{phi-p})]. However,
we notice the large error bar of this result since we have only
five data points.
We refer to \cite{unknownauthors} for a more systematic study of this
problem.



\begin{thebibliography}{}
\bibitem{footnote} In this
study, nonlinear effects refer to the fact that the elastic
moduli--- calculated using small-amplitude strains--- depend
nonlinearly on the external stress, and they do not refer to the
nonlinear mechanical response to large strain perturbations.
\end{thebibliography}
\end{document}